\documentclass[10pt,journal,final,finalsubmission,twocolumn]{IEEEtran}

\usepackage{cite} 
\usepackage{graphicx}
\usepackage{algpseudocode}
\usepackage{epsfig,epstopdf}
\usepackage{subfigure}
\usepackage{url}
\usepackage{amssymb,amsmath,amsbsy,amsfonts}
\usepackage{expdlist}
\usepackage{color}
\usepackage{rotating}
\usepackage{multirow} 
\usepackage{tabularx}
\usepackage{pdflscape}
\usepackage{booktabs}
\DeclareGraphicsExtensions{.jpg,.pdf,.png,.gif,.eps}
\graphicspath{{./Figures/}}

\usepackage{latexsym}
\usepackage{amsmath}
\usepackage{amssymb}
\usepackage{psfrag}
\usepackage{color}
\usepackage{cite}
\usepackage{times}
\usepackage{setspace}
\usepackage{multirow}
\usepackage{url}
\usepackage{hyperref}
\usepackage{bbm}
\usepackage{epstopdf}
\usepackage{graphicx,dblfloatfix}

\def\T{{\mathbf{T}}}
\def\E{{\mathbb{E}}}

\def\pe{{\epsilon}}
\def\C{{\mathcal{C}}}
\def\L{{\mathcal{L}}}

\usepackage{mathtools}
\DeclarePairedDelimiter{\ceil}{\lceil}{\rceil}

\begin{document}
%
\title{Improving the Finite-Length Performance of Spatially Coupled LDPC Codes by Connecting Multiple Code Chains}

%
%

\author{\authorblockN{Pablo M. Olmos \IEEEmembership{Member, IEEE}\thanks{This work was partially supported by NSF grant CCF11-61754, by TELUS Corporation Canada, and by Spanish government grants MEC TEC2009-14504-C02-01, TEC2012-38800-C03-01, and Consolider-Ingenio 2010 CSD2008-00010.}}\thanks{P. M. Olmos is with the Signal Theory \& Communications Department, Universidad Carlos III de Madrid.  E-mail: {\tt olmos@tsc.uc3m.es}}, \and
\authorblockN{David G. M. Mitchell\thanks{David G. M. Mitchell and Daniel J. Costello,  Jr. are with the Electrical Engineering Department, University of Notre Dame, Indiana, USA.  E-mail: {\tt \{david.mitchell, costello.2\}@nd.edu}}, \IEEEmembership{Member, IEEE},} \and
\authorblockN{Dmitri Truhachev\thanks{Dmitri Truhachev is with the ECE Dept., Dalhousie University, Halifax, Canada.  E-mail: {\tt dmitry@dal.ca}}, \IEEEmembership{Member, IEEE}}, \and \authorblockN{Daniel J. Costello,  Jr.,
\IEEEmembership{Life Fellow, IEEE}} 
\thanks{This work was presented in part at the IEEE Information Theory Workshop, Sevilla, Spain, Sept. 2013.}}

\markboth{Submitted to IEEE Transactions on Information Theory, ~February~2014}{ }

\maketitle

\begin{abstract}
In this paper, we analyze the finite-length performance of codes on graphs constructed by connecting spatially coupled low-density parity-check (SC-LDPC) code chains.   Successive (peeling) decoding  is considered for the binary erasure channel (BEC). The evolution of the undecoded portion of the  bipartite graph remaining after each  iteration is analyzed as a dynamical system. When connecting short SC-LDPC chains, we show that, in addition to superior iterative decoding thresholds, connected chain ensembles  have better finite-length performance than single chain ensembles of the same rate and length. In addition, we present a novel encoding/transmission scheme to improve the performance of a system using long SC-LDPC chains, where, instead of  transmitting   codewords corresponding to a single SC-LDPC chain independently, we connect consecutive chains in a multi-layer format to form a connected chain ensemble. We refer to such a transmission scheme to as continuous chain (CC) transmission of SC-LDPC codes. We show that CC  transmission can be implemented  with no significant increase in  encoding/decoding complexity or decoding delay with respect a system using a single SC-LDPC code chain  for encoding.
\end{abstract}

\section{Introduction}

Low-density parity-check (LDPC) block codes, invented by Gallager in the 1960's~\cite{gal63} and later rediscovered in the 1990's \cite{mac99, mac98}, have attracted a  great deal of interest due to their potential for near-capacity   performance. However, the iterative decoding techniques generally employed for LDPC decoding are suboptimal compared to optimal maximum a posteriori probability (MAP) decoding, which is prohibitively complex for the  typical operational lengths of LDPC codes. As a result, the iterative belief propagation (BP) decoding limits (thresholds) of LDPC block codes are worse than their MAP decoding thresholds \cite{Urbanke08}. 

In contrast, it has recently been proven that spatially coupled
low-density parity-check (SC-LDPC) codes can be designed to achieve
capacity over general binary-input memoryless output-symmetric (BMS)
channels under iterative BP decoding \cite{KudekarBMSIT}.  The reason for this behavior is the spatial graph coupling that defines the structure of an SC-LDPC code:  the Tanner graph of a block code with $M$ variable nodes is duplicated $L$ times to produce a sequence of identical graphs; following this, the neighboring copies are then connected to form a chain by redirecting (spreading) certain edges.  The resulting graph  has a so-called ``structured irregularity'', where parity check nodes located at both ends of the chain are connected to a smaller number of variable nodes than those in the middle \cite{lscz10}. As a result, the nodes at the ends of the graph form strong sub-codes and the resulting reliable information generated there during BP decoding propagates through the chain toward the center. It has been shown that, for sufficiently large $L$, the BP decoding thresholds of $(l,r)$-regular SC-LDPC codes coincide with the MAP decoding thresholds of the underlying $(l,r)$-regular LDPC block codes \cite{lscz10} \cite{kru11}, which are known to approach capacity with increasing graph density. As a result, the principle of spatial graph coupling has attracted significant attention and has been successfully applied in many other areas of communications and signal processing~\cite{kp10},\cite{Truhachev12},\cite{ttk11}.

Spatial graph coupling need not be limited to the connection of graphs to form a single chain. In \cite{Truhachev13, TruhachevICC12}, more general ensembles were proposed that are constructed by connecting together several individual SC-LDPC code chains. It was 
demonstrated that, by optimizing the connection points, the lengths of the connected chains, and their densities,
ensembles with improved decoding thresholds can be constructed. A particularly interesting example of such a construction is the loop ensemble~\cite{Truhachev13, TruhachevICC12}. For short chain lengths, \emph{e.g.}, $L$ between $5$ and $20$, the loop ensemble constructed by connecting two $(l,r)$-regular SC-LDPC code ensembles has significantly better BP thresholds than  single component chains of the same rate and code length. 
However, for long chain lengths, no constructions have yet been  discovered for ensembles created by connecting multiple  SC-LDPC chains that are able to outperform a single chain of the same rate and code length. 

In contrast to  most of the performance analysis of SC-LDPC codes that typically focuses on the asymptotic (in code length) behavior of code ensembles (see, \emph{e.g.}, \cite{KudekarBMSIT}\cite{kru11}\cite{lscz10},\cite{lmfc10b}), in this paper we are interested in the \emph{finite-length} performance of SC-LDPC codes. Analysis of the finite-length performance of  LDPC codes in the waterfall region of the bit-error rate (BER) curve is typically performed for the case of transmission over the binary erasure channel (BEC), where scaling laws relating the finite-length code performance and the LDPC code parameters can be analytically computed \cite{Urbanke09, Nozaki12}. It has also been observed that such scaling laws are essentially valid across all BMS channels \cite{Ezri07}. For the BEC, we consider an equivalent formulation to BP called peeling decoding (PD) \cite{Urbanke08-2}. PD iteratively removes variable nodes with known (decoded) values  from the Tanner graph. This yields a sequence of graphs whose statistics define both the asymptotic and  finite-length properties of the code \cite{Luby01}. An estimate of the PD error probability throughout the decoding process can be obtained based on the expected or mean evolution of the number of degree-one check nodes, 
 as well as the variance around this evolution \cite{Urbanke09}.  Indeed, the PD error probability is dominated by local minima in the expected evolution of the number of degree-one check nodes \cite{Amraoui05}. We refer to these minima as critical points.

Scaling laws for single  SC-LDPC  chains  have been recently proposed  for the BEC in \cite{Olmos13,Olmos13-3}, for large $L$ values,  and in \cite{Olmos13-2} for small values of $L$. In this paper, we investigate the finite-length properties of a variety of single and connected SC-LDPC code chains. 
 An important result  is that the finite-length performance of short and long SC-LDPC chains is governed by scaling laws with very different characteristics. For short chains, where the code rate is lower, the expected evolution of degree-one check nodes displays only a single critical point, similar to regular LDPC codes \cite{Amraoui05}; however, for (higher rate) long chains,   critical points span the entire expected evolution curve and thus the decoder might fail with roughly uniform probability at any point in the process. From this point of view, shorter chains can be considered to be more robust against decoding failures. This result has not yet been presented in the literature and it is the first main contribution of this paper. Note the important trade-off from the design perspective, already pointed out in \cite{Iyengar11,Olmos11-3}: while channel capacity can only be achieved in the limit of large $L$, for a given finite code length $n=ML$ and a minimum acceptable design rate, the performance of a SC-LDPC code with shorter chain length $L$ and larger $M$ may be better than for a code with longer $L$ and smaller $M$, even though the threshold of the 
 latter code is closer to capacity. 

Next, we build on the results of \cite{Truhachev13, TruhachevICC12,Olmos13-2} and investigate the performance of connected SC-LDPC code chains using PD.  In particular, we study the finite-length properties of the connected ``loop'' ensemble. We show that,  for short chain lengths, not only the loop has  a threshold  closer to capacity than the single chain ensemble, also it has better finite-length performance compared to the single chain ensemble of the same rate and code length.  The expected evolution of degree-one check nodes is analytically computed for both ensembles and we show that, in each case, the performance is dominated by a single critical point. Besides, we find that
there is little or no improvement with this construction for longer chains. To provide a better understanding of the decoding dynamics of connected chains, we revisit the role of those positions at which two or more chains are linked. To this end, we first consider a modified chain ensemble, consisting of two regions with higher degree variable nodes rather than the typical construction with regions of lower degree check nodes. We show that this method of constructing an SC-LDPC code chain has important weaknesses and the beneficial effect of spatial coupling  is lessened.  This behavior can then be used to explain the limitations of the loop ensemble when constructed using long SC-LDPC chains and to give guidelines on how to construct code ensembles with improved performance in this regime. 

Based on this new understanding of the decoding dynamics of short and long chains, the next contribution of this paper is to present a new transmission technique using SC-LDPC code chains that improves the finite-length error rate performance, in particular when long chains are used. In the case of a single terminated SC-LDPC  code chain, the information is encoded into independent codewords corresponding to unconnected consecutive SC-LDPC chains. We  propose to link consecutive chains by creating a dependence between them, thus forming a special pattern corresponding to an  SC-LDPC code based on connected chains.  This coding strategy resembles block Markov superposition encoding \cite{Kramer05}, first proposed by Cover and El Gamal in \cite{Cover79} for the relay channel.  We refer to such an encoding/transmission scheme as continuous chain (CC) transmission. We also propose a way to exploit the two strong sub-codes at both ends of the individual  SC-LDPC chains so that the connected chain results in  a stronger  code of the same rate. The underlying principle is to generate reliable information at various points in the graph - effectively breaking long chains into better protected shorter chains. CC transmission is shown to be feasible for both encoding and decoding.  In particular, CC transmission only requires some additional memory in the encoding process and, to effectively implement  a windowed decoder \cite{lscz10,Iyengar11}, a different order in the sequence of transmitted bits compared to transmitting unconnected chains. 
%
%
To the best of our knowledge, this is a  new concept in the field of SC-LDPC codes. The final contribution of this paper is to propose a number of designs for SC-LDPC ensembles based on connected chains and optimized for CC transmission. Simulation results for  the BEC and the binary input additive white Gaussian noise (BIAWGN) channel show that the  finite-length error rate performance  improves significantly, even for extremely long code lengths, \emph{e.g.},   $M=2000$ bits per position and a chain length of $L=50$.

In Section \ref{singlechain} we review the construction of a single SC-LDPC code chain and we analyze its finite-length performance. In Section \ref{modified_chain} we introduce a modification to this ensemble, in which strong sub-codes composed of low-degree check nodes are replaced by groups of variable nodes of higher degree. Studying the characteristics of this modified ensemble provides an explanation for the performance improvement obtained when connecting SC-LDPC chains, particularly  for the loop construction  in Section \ref{loop}. In Section \ref{SectionV} we present the CC transmission scheme, propose SC-LDPC code constructions based on connected chains that are well suited for CC transmission, and give simulation results that illustrate its benefits. Finally, we provide some concluding remarks in Section \ref{future},  including potential future lines of research  on this topic.

\section{The  SC-LDPC Single Chain Ensemble}\label{singlechain}


In this section we review the construction of a  SC-LDPC code chain.  We use the random construction of SC-LDPC ensembles presented in \cite{Olmos13-3} and the BEC for most of our analysis. As discussed in the introduction, a finite-length analysis of LDPC codes is feasible for the BEC and, as explained in \cite{Olmos13-3}, using the random construction greatly simplifies the analysis. A particularly interesting class of codes can be constructed by means of protographs \cite{tho03}, with important practical advantages \cite{Lentmaier10-2,Mitchell11-2}; however, these constructions, which can be considered as  a subclass of multi-edge type LDPC code \cite{Urbanke08-2}, significantly complicate the analysis. Nevertheless, the results on connected chains that we derive in the paper can be directly applied to the design of protograph-based SC-LDPC code ensembles.

We start by describing the $(l,r)$-regular SC-LDPC single chain  ensemble. Consider a chain of $L$ uncoupled $(l,r)$-regular  LDPC block codes of length $M$ bits that occupy $L$ consecutive positions. By $l$ ($r$) we denote the variable (check) node degree of the regular LDPC block code. Each block code has $M$ variable nodes of degree $l$ and $\frac{l}{r}M$ check nodes of degree $r$.  We generate a SC-LDPC code ensemble by spreading $l-1$ edges per variable node to check nodes associated with nearby codes in the chain. 

\begin{figure}[h]
\centering \includegraphics[scale=1.85]{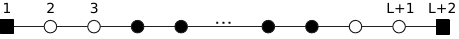}
\caption{A  single SC-LDPC $\mathcal{C}(3,6,L)$ chain. White nodes indicate positions where variable nodes are connected to check nodes of degree less than $r$. Square nodes indicate positions with no variable nodes. }\label{Fig:parallel}
\end{figure}

Without loss of generality, we demonstrate the construction of an SC-LDPC code ensemble based on an uncoupled $(3,6)$-regular LDPC block code. The $\mathcal{C}(3,6,L)$ single chain ensemble that we consider  is generated as follows. We place $L+2$ groups of $M/2$ check nodes at positions $u=1,2,\ldots,L+2$ and $L$ groups of $M$ variable nodes of degree three at positions $u=2,\ldots,L+1$.  The $3$ edges of each variable node located at position $u$ are connected to  check nodes at positions $u-1$, $u$, and $u+1$.  These check nodes are chosen uniformly at random from the $M/2$ check nodes at each position.
%
%
\textcolor{black}{At the middle positions $4,\ldots,L-1$, each check node has $6$ \emph{sockets} where edges coming from variable nodes can be connected. So in total, 
the check nodes in the middle positions have $M/2\times 6$ sockets  and there are  $3\times M$ incoming edges from variable nodes. Thus, after edge spreading all check nodes in the middle positions are of degree $6$, since there are no free sockets left.} 

The $M/2$ check nodes located at positions $1$ and $L+2$  have $M$ incoming edges from variable nodes at positions $2$ and $L+1$, respectively, \textcolor{black}{and thus  the number of sockets per check node is only $2$, so all check nodes have degree  $2$ after edge spreading.}
Similarly, the $M/2$ check nodes at positions $2$ and $L+1$  have $2M$ incoming edges, so  the check node degree at these two positions is $4$. Overall, the coupled code occupies $L+2$ positions, labeled from $1$ to $L+2$, but  there are variable nodes only in positions $2$ to $L+1$. 
Any graph generated using  this procedure  defines a code sample from  the $\mathcal{C}(3,6,L)$ single chain ensemble.  Hereafter, we graphically represent a single chain as shown in Fig.~\ref{Fig:parallel}, where the round dots indicate positions with variable nodes. White nodes indicate positions where variable nodes are connected to check nodes of degree less than $r$, \emph{i.e.}, they represent regions with better protection as a result of the lower-degree check nodes. Square nodes indicate positions with no variable nodes.

The $\mathcal{C}(3,6,L)$ ensemble can be represented in a compact way using an $(L+2)\times (L+2)$ binary \emph{connectivity matrix} $\T$. $T_{uv}=1$ means that the $M$ variable nodes at position $v$ are connected with one edge to a randomly chosen check node at position $u$. Let $d_{c,u}$ denote the sum of elements in the $u$-th row of $\T$ and $d_{v,u}$ the sum of the elements in the $u$-th column. The degree of the  check nodes at position $u$ is  given by $\frac{r}{l}d_{c,u}$, and we get $d_{v,u}=0$ for the positions with no variables nodes, i.e, the end positions. For example, the $10\times 10$ $\T$ matrix for a single $(3,6)$-regular chain with $L=8$ is
\begin{align}\label{Tmatrix}
\T_{\mathcal{C}(3,6,8)}=
 \left[
                  \begin{array}{cccccccccc}
                    0\! & 1\! & 0\! & 0\! & 0\! & 0\! & 0\! & 0\! & 0\! & 0\!  \\
                    0\! & 1\! & 1\! & 0\! & 0\! & 0\! & 0\! & 0\! & 0\! & 0\!  \\
                    0\! & 1\! & 1\! & 1\! & 0\! & 0\! & 0\! & 0\! & 0\! & 0\!  \\
                    0\! & 0\! & 1\! & 1\! & 1\! & 0\! & 0\! & 0\! & 0\! & 0\!  \\
                    0\! & 0\! & 0\! & 1\! & 1\! & 1\! & 0\! & 0\! & 0\! & 0\! \\
                    0\! & 0\! & 0\! & 0\! & 1\! & 1\! & 1\! & 0\! & 0\! & 0\! \\
                    0\! & 0\! & 0\! & 0\! & 0\! & 1\! & 1\! & 1\! & 0\! & 0\!\\
                    0\! & 0\! & 0\! & 0\! & 0\! & 0\! & 1\! & 1\! & 1\! & 0\!  \\
                    0\! & 0\! & 0\! & 0\! & 0\! & 0\! & 0\! & 1\! & 1\! & 0\!\\
                    0\! & 0\! & 0\! & 0\! & 0\! & 0\! & 0\! & 0\! & 1\! & 0\!  \\
                  \end{array}
\right].
\end{align}
Throughout the paper,  SC-LDPC ensembles will be presented using a $D\times D$ connectivity matrix $\T$. Any SC-LDPC code based on an $(l,r)$-regular uncoupled LDPC code with connectivity matrix $\T$ is  generated as follows. At  position $u=1, \ldots, D$, we place $M$ variable nodes of degree $d_{v,u}$ and $\frac{l}{r}M$ check nodes with \textcolor{black}{ $\frac{r}{l}d_{c,u}$ free sockets each}. Then we randomly connect the $d_{v,u}$ edges of each variable node at position $u$ to check nodes at positions given by the ones of the $u$-th colum of $\T$. The design rate of the SC-LDPC ensemble can be computed from the $\T$ matrix. More specifically, the number of check nodes in the Tanner graph is given by $D\frac{l}{r}M$ and the code length is  $M\sum_{u=1}^{D} \mathbbm{1}[d_{v,u}>0]$, where $\mathbbm{1}[d_{v,u}>0]$ equals one if $d_{v,u}>0$ and zero otherwise. For example, for the $\mathcal{C}(3,6,L)$ ensemble, $D=L+2$, $\frac{r}{l}=2$ and the design rate  is
\begin{align}\label{designrate}
\mathtt{r}_{\mathcal{C}(3,6,L)}&=1-\frac{(L+2)\frac{M}{2}}{M\sum_{u=1}^{L+2} \mathbbm{1}[d_{v,u}>0]}=\frac{1}{2}-\frac{1}{L}.
\end{align}

\subsection{Degree Distribution}
The concept of code degree distribution (DD) for uncoupled LDPC block codes can be generalized to SC-LDPC codes using the $\T$ matrix. Recall that the right degree of a given edge is defined as the degree of the check it is connected to \cite{Luby01}. Let $R_{j,u}$ be the number of edges with right degree $j$ at position $u$.
Since each variable node has at most one edge connected to a particular position, it is sufficient to know the  number $V_u$ of variable nodes per position. Given the connectivity matrix $\T$, the DD of the SC-LPDC code is, for $u=1,\ldots,D$,
\begin{align}\label{ExDD1}
V_u&=\left\{\begin{array}{cc}
M & d_{v,u}>0\\
0 & \text{ otherwise }
\end{array}
\right. ,\\
R_{j,u}&=\left\{
\begin{array}{cc}
d_{c,u}M & j=\frac{r}{l}d_{c,u}\\
0 & \text{ otherwise }
\end{array}
\right. .
\end{align}


\subsection{Peeling Decoding and expected graph evolution}\label{expected}
Now we consider transmission over the BEC  and  PD at the receiver \cite{Urbanke08-2}. We start the PD algorithm by removing from the graph all the variable nodes and edges associated with non-erased symbols,  plus any disconnected check nodes. At each iteration,  PD looks for a degree-one check node, which is removed along with the variable node it is connected to. In \cite{Luby01},  it was shown that if we apply PD to an LDPC code, the sequence of graphs follows a typical path or expected evolution.  Based on the graph covariance evolution  at those points where the expected evolution of the number of degree-one check nodes presents a local minimum (critical points), \emph{scaling laws} (SLs) for PD were proposed to predict  the finite-length performance of LDPC code  ensembles in the waterfall region  in \cite{Urbanke09}. To illustrate the main differences in the decoding of short and long SC-LDPC code chains, if suffices to observe the  ensemble expected graph evolution over the PD  process. The expressions provided are valid for computing the expected graph evolution of any SC-LDPC code  constructed according a given $\T$ matrix.

Assume a BEC with erasure probability $\epsilon$. After initializing PD  (iteration $\ell=0$) by reducing the graph according to the correctly received information,  the expected DD at position $u=1,2,\ldots,D$ is
\begin{align}\label{BEC1}
\E[V_u(\ell=0)]&=\epsilon M \mathbbm{1}[d_{v,u}>0], \\\label{BEC2}
\E[R_{j,u}(\ell=0)]&=j\frac{l}{r}M\binom{\frac{r}{l}d_{c,u}}{j}\epsilon^{j}(1-\epsilon)^{(\frac{r}{l}d_{c,u}-j)},
\end{align}
for $j\leq \frac{r}{l}d_{c,u}$, where \eqref{BEC2} comes from the fact that each one of the $\frac{r}{l}d_{c,u}$ edges of any check node at position $u$ is independently removed with probability $\epsilon$.  Let $V_u(\ell)$ and $R_{j,u}(\ell)$ be the (random) DDs at position $u$ after $\ell$ iterations. Then the normalized (random) DDs at normalized time $\tau$ can be described by
\begin{align}\label{norm}
\tau\doteq\frac{\ell}{\epsilon M}, \qquad r_{j,u}(\tau)\doteq\frac{R_{j,u}(\ell)}{\epsilon M}, \qquad v_{u}(\tau)\doteq\frac{V_u(\ell)}{\epsilon M},
\end{align} 
for $u=1,2,\ldots,D$ and $j=1,2,\ldots,r$. Note that, on average, the total number of PD iterations is $\epsilon M\sum_{u} \mathbbm{1}[d_{v,u}>0]$, and thus $\tau\in[0,\sum_{u} \mathbbm{1}[d_{v,u}>0]]$. For the $\mathcal{C}(l,r,L)$ ensemble, $\tau\in[0,L]$.

Let $\hat{r}_{j,u}(\tau)$ and $\hat{v}_{u}(\tau)$ denote the mean value of  the random processes $r_{j,u}(\tau)$ and $v_{u}(\tau)$ in \eqref{norm}. As shown in \cite{Luby01},  $\hat{r}_{j,u}(\tau)$ and $\hat{v}_{u}(\tau)$ can be computed as the solution to the following   system of differential equations:
\begin{align}\label{system1}
\frac{\partial \hat{r}_{j,u}(\tau)}{\partial \tau}&=\E[R_{j,u}(\ell+1)-R_{j,u}(\ell)\Big|\hat{r}_{q,m}(\ell),\hat{v}_{m}(\ell) \; \forall m,q]\hspace{-2mm}\\\label{system2}
\frac{\partial \hat{v}_{u}(\tau)}{\partial \tau}&=\E[V_{u}(\ell+1)-V_{u}(\ell)\Big|\hat{r}_{q,m}(\ell),\hat{v}_{m}(\ell) \; \forall m,q]
\end{align}
for $u=1,2,\ldots,L+2$ and $j=1,2,\ldots,r$. Note that the right hand-side of \eqref{system1} represents the expected change in the number of check nodes of degree $j$ at position $u$ after iteration $\ell+1$ if the DD at iteration $\ell$ is exactly at its mean. Further, the solution of \eqref{system1} and \eqref{system2} is unique and, with probability $1-\mathcal{O}(\text{e}^{-\sqrt{M}})$, 
 any particular realization of the normalized DD in (\ref{norm}) deviates from its mean by a factor of less than $M^{-1/6}$ for the initial conditions $\hat{r}_{j,u}(0)=\E[R_{j,u}(\ell=0)]/\epsilon M$ and $ \hat{v}_{u}(0)=\E[V_u(\ell=0)]/\epsilon M$. The computation of the expectations in \eqref{system1} and \eqref{system2} can be found in Appendix \ref{A1}.  
Consequently,  the ensemble  threshold under BP decoding  can be computed as  the maximum $\pe$ for which the mean of the normalized number of degree-one check nodes in the graph, given by
\begin{align}\label{r1mean}
\hat{r}_{1}(\tau)\doteq\sum_{m=1}^{D}\hat{r}_{1,m}(\tau),
\end{align}
is positive for any $\tau\in[0, \sum_{i=1}^{D} \mathbbm{1}[d_{v,u}>0]]$. Based on (\ref{r1mean}), we can say that $\hat{r}_{1}(\tau)$ is the mean of the random process $r_{1}(\tau)=\sum_{m=1}^{D}r_{1,m}(\tau)$.


\begin{figure}[h]
\centering \includegraphics[scale=0.5]{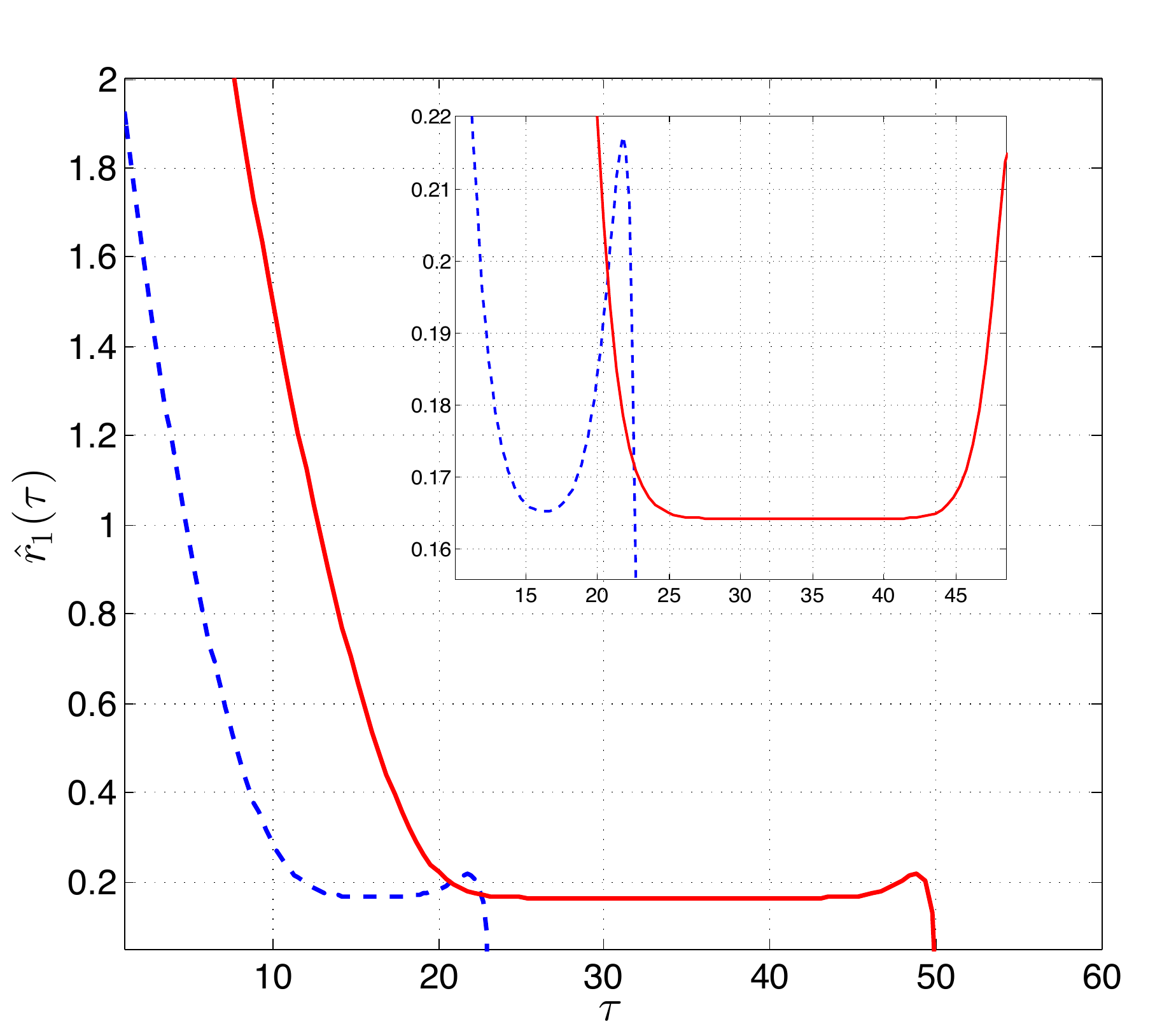}
\caption{Solution to $\hat{r}_{1}(\tau)$ for the ensembles $\mathcal{C}(3,6,25)$ (dashed line) and $\mathcal{C}(3,6,50)$ (solid line) for $\epsilon=0.45$. A magnification is shown in the upper-right corner of the figure.}\label{Fig: r1chain}
\end{figure}

\subsection{Scaling behavior}\label{scaling}

In Fig.~\ref{Fig: r1chain}, we plot $\hat{r}_{1}(\tau)$ for the ensembles $\mathcal{C}(3,6,25)$ (dashed line) and $\mathcal{C}(3,6,50)$ (solid line) and $\epsilon=0.45$. After magnification, we observe that $\hat{r}_{1}(\tau)$ for the 
$\mathcal{C}(3,6,25)$ ensemble presents a unique local minima at $\tau^*\approx16$. As shown in \cite{Olmos13-2}, the ensemble average block error probability for short SC-LDPC code chains with a single critical point at a given time $\tau^*$ is well predicted by the cumulative probability distribution of the process $r_1(\tau)$ at that time. For sufficiently large $M$,  $r_1(\tau)$ is approximately Gaussian distributed \cite{Urbanke09} and thus the block error probability can be estimated as follows:
\begin{align}\label{shortchain}
P_{\mathcal{C}(l,r,L)}\approx \mathcal{Q}\left(\frac{\sqrt{M}\Delta_{\epsilon}}{\alpha}\right),
\end{align}  
where $\Delta_{\epsilon}=(\epsilon_{\mathcal{C}(3,6,25)}-\epsilon)$,  $\epsilon_{\mathcal{C}(3,6,25)}$ is the $\mathcal{C}(3,6,25)$ ensemble's BP threshold and $\alpha=\alpha(l,r)$ is  the ratio of  the standard deviation of the process $r_1(\tau)$ to its mean $\hat{r}_{1}(\tau)$, computed at the critical point $\tau^*$.

In contrast, as demonstrated for the $\mathcal{C}(3,6,50)$ ensemble in Fig. \ref{Fig: r1chain}, for longer chains $r_1(\tau)$ does not display a single critical point, but rather a critical phase  in which it remains constant. During this phase, the decoder might fail roughly at any point with uniform probability.  As explained in \cite{Olmos13}, the dynamics of the decoding process are governed by  so-called wave decoding \cite{kru11,Olmos11-3}. After positions in the middle of the chain have run out of degree-one check nodes, which happens at $\tau\approx25$  in Fig. \ref{Fig: r1chain}, recovered information from the boundary positions propagates at constant speed through the chain to the middle positions. 
These dynamics explain the constant value of $\hat{r}_1(\tau)$  during  the so-called steady-state phase. In \cite{Olmos13, Olmos13-3}, the block error probability in the steady-state phase is  estimated as follows:
\begin{align}\label{longchain}
P_{\mathcal{C}(l,r,L)}&\approx 1-\exp\left(-\frac{\epsilon L \overline{y}}{\mu_0(M,\epsilon,l,r)}\right),
\end{align}
where $ \overline{y}=\overline{y}(l,r,\epsilon)$ is a quantity that measures how long the decoder  remains in the critical phase and $\mu_0$ is the average survival time of the $r_1(\tau)$ process during the critical phase before crossing the zero level, which implies a decoding error. As shown in \cite{Olmos13-3}, $ \overline{y}$ can be bounded using  BP density evolution for the $(l,r)$-regular LDPC code ensemble. Also, $\mu_0$ can be expressed as follows:
\begin{align}\label{mu}
\mu_0(l,r,M,\epsilon)\approx\frac{\sqrt{2\pi}}{\theta}\int_{0}^{\frac{\sqrt{M}\Delta_\epsilon}{\alpha}}\Phi(z)\text{e}^{\frac{1}{2}z^2}\text{d}z,
\end{align}
where $\Phi(z)$ is the  c.d.f. of the standard Gaussian distribution, $\mathcal{N}(0,1)$, and $\theta=\theta(l,r)$ a parameter related to the exponential decay of the covariance of  $r_1(\tau)$ with time, i.e., $\text{CoV}[r_1(\tau), r_1(\zeta)]\propto \exp(-\theta |\zeta-\tau|)$  
(see \cite{Olmos13-3} for further details). 

\begin{figure}[h]
\centering \includegraphics[scale=0.45]{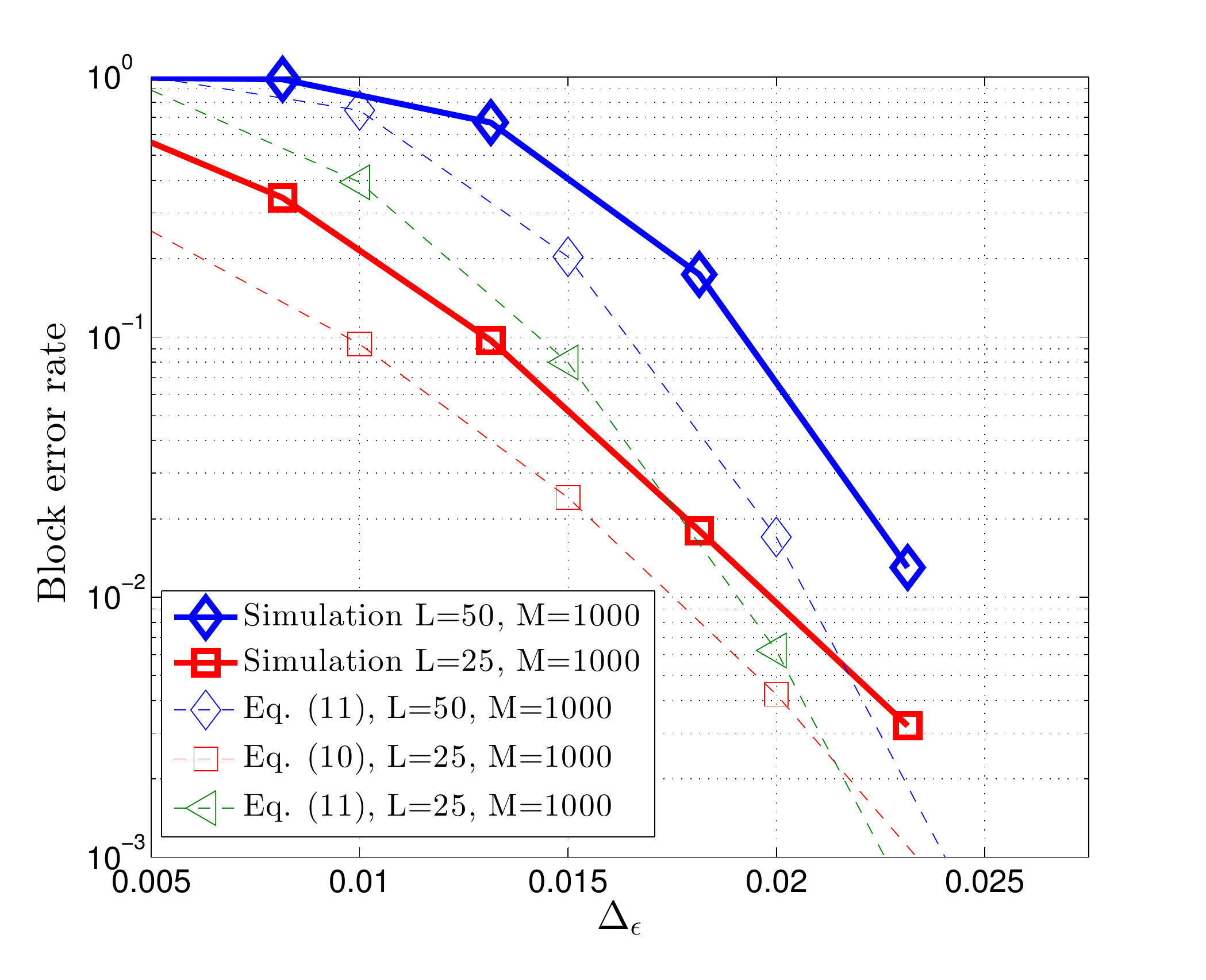}
\caption{Simulated block error rate (solid lines) of the ensembles  $\epsilon_{\mathcal{C}(3,6,50)}$ and $\epsilon_{\mathcal{C}(3,6,25)}$ with $M=1000$ bits per position as a function of $\Delta_\epsilon$. Also included is the block error rate estimate obtained using \eqref{shortchain} and \eqref{longchain} (dashed lines).}\label{WERchains}
\end{figure}

 The scaling law for short SC-LDPC chains in (\ref{shortchain}) can be considered an extreme case of \eqref{longchain}, where wave-like decoding is reduced to a single time instant.
 For the same distance $\Delta_\epsilon$ to their respective thresholds and the same number $M$ of bits per position, the process $r_1(\tau)$ has similar statistical properties at the critical point/phase. For instance, in Fig.~\ref{Fig: r1chain} we  observe that the mean height at the respective minima is just slightly higher for the $\mathcal{C}(3,6,25)$ case, and by simulation we can confirm that, at these points, the  standard deviation of $r_1(\tau)$ is similar in both cases. However, the dynamics of decoding long chains require the process $r_1(\tau)$ to survive a period of time that grows linearly with $L$ and during which its mean is very low. Thus the decoding process is more susceptible to errors. 

 In Fig.~\ref{WERchains}, we plot the simulated block error rate (solid lines) of the ensembles  $\epsilon_{\mathcal{C}(3,6,50)}$ and $\epsilon_{\mathcal{C}(3,6,25)}$ with $M=1000$ bits per position. Despite the fact that the ensembles have different  rates, the performance comparison is fair, since it is based on the distance to their respective thresholds. We also include the block error rate estimate  obtained using \eqref{shortchain} and \eqref{longchain} with the following values: $\alpha_{\mathcal{C}(3,6,25)}\approx0.22$, $\alpha_{\mathcal{C}(3,6,50)}\approx0.17$, and $\theta=0.59$\footnote{The difference between the simulated performance and that estimated using scaling laws is mainly because, when using a scaling law such as \eqref{shortchain} or \eqref{longchain}, we neglect some second order effects that are important for low values of $M$ \cite{Urbanke09,Olmos13-2,Olmos13}. For example, in \eqref{shortchain} we ignore the fact that, with non-zero probability, $r_1(\tau)$ can equal zero for some $\tau\neq\tau^*$.}. In \eqref{longchain}, $\overline{y}$  can be computed as shown in \cite{Olmos13}.  We have also included the block error rate estimate for $L=25$ using the long-chain model in \eqref{longchain}, which demonstrates the accuracy of \eqref{shortchain} and \eqref{longchain} when used for SC-LDPC code ensembles in the correct regime.

Two important remarks can be made based on Fig.~\ref{WERchains}. First, at the same distance to their respective  thresholds, the ensemble $\mathcal{C}(3,6,25)$ provides a significantly lower block error rate. This fact can be observed in both the simulation curves and the estimated block error rate curves. Second, this gain is not just explained by the smaller $L$ value in \eqref{longchain}. There is a change in the dynamics of the decoding process and thus  qualitative improvement in the robustness of the code. This can be observed in Fig. \ref{WERchains} by comparing the estimated performance for $L=25$ using \eqref{shortchain} ($\square$ marker) and \eqref{longchain} ($\lhd$ marker). This result is of significant practical importance and has not yet  been discussed in the literature. In  Section \ref{SectionV}, we  exploit this result to improve the finite-length performance of systems based on long SC-LDPC code chains.


\section{A modified SC-LDPC Single Chain  ensemble}\label{modified_chain}

In this section, we introduce a modification of the chain ensemble $\C(3,6,L)$ presented above to illustrate the effect of introducing a structured irregularity into the SC-LDPC code chain in a different way. The analysis of this modified code chain allows us to understand, analyze, and design SC-LDPC code ensembles based on connected chains with improved performance, as we demonstrate in Sections \ref{loop} and \ref{SectionV}. 


\begin{figure}[h]
\centering \includegraphics[scale=3]{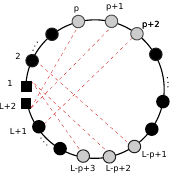}
\caption{Graphical representation of the $\widetilde{\mathcal{C}}(3,6,L,p)$ ensemble. Dashed lines represent the additional edges added with respect to the $\C(3,6,L)$ ensemble and gray dots indicate positions with variable nodes of degree four.}\label{Fig:close_chain}
\end{figure}

We again consider the $\C(3,6,L)$ ensemble as an illustrative example.  Let $p\in\{2,\ldots,\ceil{L/2}\}$  be a position in the code chain that contains variable nodes.  We define the ensemble $\widetilde{\mathcal{C}}(3,6,L,p)$ as follows: given the $\C(3,6,L)$ ensemble, we add one additional edge from each variable node at positions $p$ and $p+1$ to a random and uniformly chosen check node at position $L+2$
and one additional edge between each variable node at position $p+2$ to a check node at position $L+1$.
We  proceed symmetrically by adding edges in a similar fashion, connecting variable nodes at positions $L-p+3,L-p+2$, and $L-p+1$ to check nodes at the left end of the chain, \emph{i.e.}, positions $1$ and $2$. These connections are chosen so that we do not increase the check node degree beyond $6$. For example, the connectivity matrix $\T$ of the $\widetilde{\mathcal{C}}(3,6,8,p)$ ensemble for $p=3$ is 
\begin{align}\label{Tmatrix2}
\T_{\widetilde{\mathcal{C}}(3,6,8,3)}=
 \left[
                  \begin{array}{cccccccccc}
                    0\! & 1\! & 0\! & 0\! & 0\! & 0\! &\textcolor{red}{\mathbf{1}}\! & \textcolor{red}{\mathbf{1}}\! & 0\! & 0\!  \\
                    0\! & 1\! & 1\! & 0\! & 0\! & \textcolor{red}{\mathbf{1}}\! & 0\! & 0\! & 0\! & 0\!  \\
                    0\! & 1\! & 1\! & 1\! & 0\! & 0\! & 0\! & 0\! & 0\! & 0\!  \\
                    0\! & 0\! & 1\! & 1\! & 1\! & 0\! & 0\! & 0\! & 0\! & 0\!  \\
                    0\! & 0\! & 0\! & 1\! & 1\! & 1\! & 0\! & 0\! & 0\! & 0\! \\
                    0\! & 0\! & 0\! & 0\! & 1\! & 1\! & 1\! & 0\! & 0\! & 0\! \\
                    0\! & 0\! & 0\! & 0\! & 0\! & 1\! & 1\! & 1\! & 0\! & 0\!\\
                    0\! & 0\! & 0\! & 0\! & 0\! & 0\! & 1\! & 1\! & 1\! & 0\!  \\
                    0\! & 0\! & 0\! & 0\! & \textcolor{red}{\mathbf{1}}\! & 0\! & 0\! & 1\! & 1\! & 0\!\\
                    0\! & 0\! & \textcolor{red}{\mathbf{1}}\! & \textcolor{red}{\mathbf{1}}\!  & 0\!  & 0\! & 0\! & 0\! & 1\! & 0\!  \\
                  \end{array}
\right],\nonumber
\end{align}
where $d_{c,u}=3$ for $u=1,2,\ldots,D=L+2$, and thus all check nodes in the graph have degree six. Hence, the graph is \emph{check regular}, as opposed to the standard \emph{variable regular} SC-LDPC code construction.

 By including these additional edges, we have \emph{closed} the chain, losing the low-degree check nodes at the ends of the $\C(3,6,L)$ graphs; however, we have created two regions with better protected symbols, where the variable nodes at three consecutive positions in the chain have degree four rather than three. In Fig.~\ref{Fig:close_chain}, we give a graphical representation of the $\widetilde{\mathcal{C}}(3,6,L,p)$ ensemble. Dashed lines represent the additional edges added with respect to the $\C(3,6,L)$ ensemble and gray dots indicate positions with variable nodes of degree four. 

\begin{figure}[h]
\centering \includegraphics[scale=0.45]{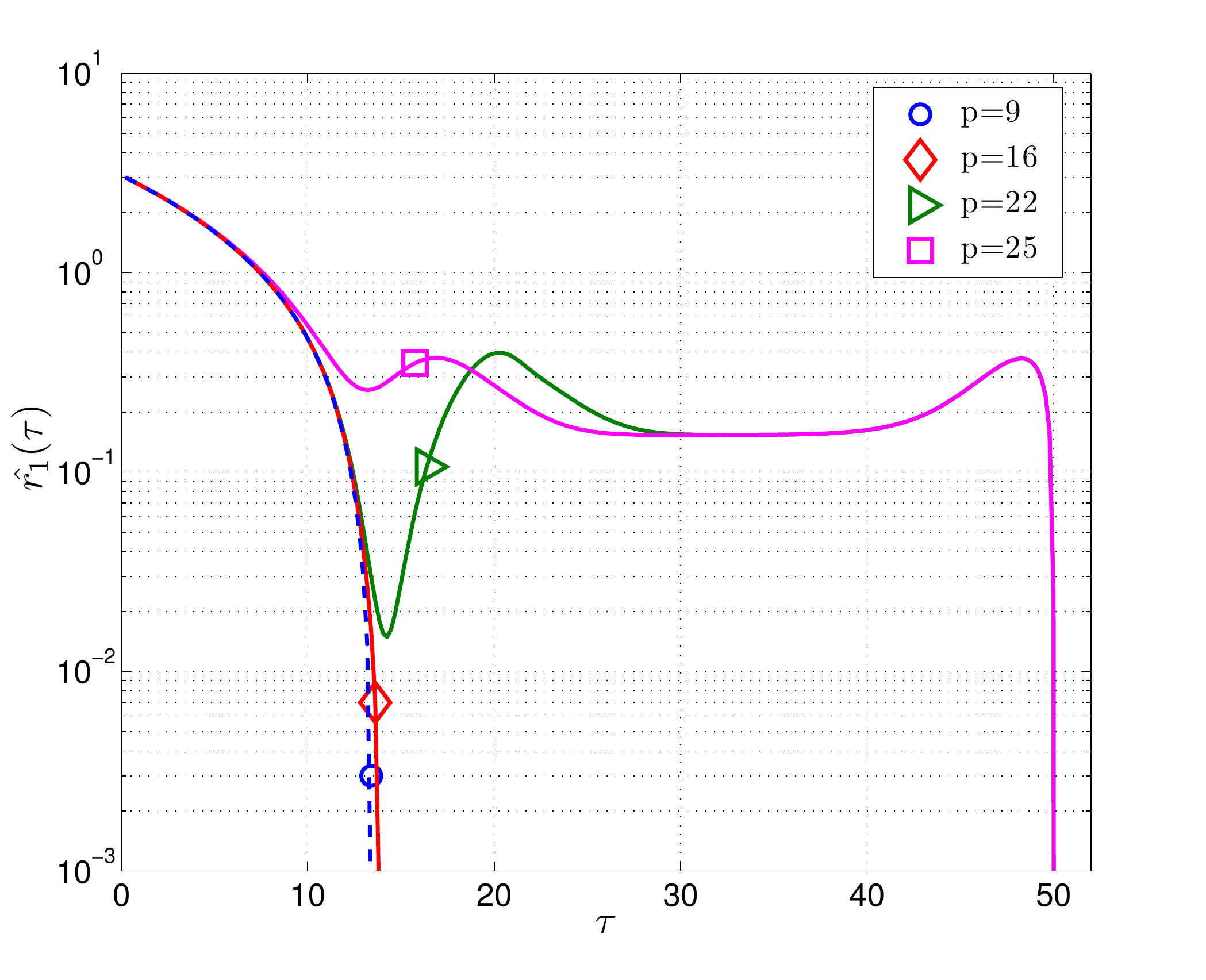}
\caption{For the ensemble $\widetilde{\mathcal{C}}(3,6,50,p)$, we represent $\hat{r}_1(\tau)$ in \eqref{r1mean} for $\epsilon=0.47$ and $p=9$, $p=16$, $p=22$ and $p=25$. }\label{Fig:close_chain_r1}
\end{figure}

The expected evolution of the normalized number of degree-one check nodes in the graph, \emph{i.e.}, $\hat{r}_1(\tau)$ in \eqref{r1mean},  for the ensemble $\widetilde{\mathcal{C}}(3,6,50,p)$ with $\epsilon=0.47$ is shown in Fig.~\ref{Fig:close_chain_r1}. We consider different values for $p$: $p=9$, $p=16$, $p=22$, and $p=25$. First, we note that the new  BP threshold $\pe_{\widetilde{\mathcal{C}}(3,6,50,p)}$ is not independent of $p$, and it is even below $0.47$ for $p=9$ and $p=16$. For small values of $p$, the two better protected regions defined by variables nodes of degree four (shown in  Fig. \ref{Fig:close_chain} by gray circles) are far away from each other. When isolated in this way, unlike the regions protected by the low degree check nodes in the standard  $\C(3,6,50)$ ensemble, they are too weak to initiate  wave-like decoding towards the interior positions of the code. As a consequence, the BP threshold is significantly decreased compared to the threshold of the standard  ensemble ($\pe_{\C(3,6,L)}\approx 0.488$). In Table~\ref{TABLE1}, we present the BP thresholds of the $\widetilde{\mathcal{C}}(3,6,50,p)$ code ensembles as a function of $p$. As we increase $p$, moving the two better protected regions closer together, the threshold quickly approaches $0.488$. The maximum threshold  is achieved for $p=25$, which corresponds to a single region of the chain where there are four consecutive positions (positions $25$, $26$, $27$, and $28$) with variable node degree larger than $3$ ($4$, $5$, $5$, and $4$, respectively). Note that, in this case, there is no threshold degradation with respect to the  $\C(3,6,50)$  ensemble. From the finite-length scaling perspective (see Fig.~\ref{Fig:close_chain_r1}), note  the presence of a critical point at $\tau\approx 14$ for the ensembles with $p=22$ and $p=25$. For the case $p=22$, this critical point dominates the block error rate; however, for $p=25$, most errors in the finite-length regime occur in the critical phase corresponding to  wave-like decoding towards the interior positions of the code. In Section~\ref{sims}, we show by simulation that the finite-length performance of the ensembles $\C(3,6,50)$ and $\widetilde{\mathcal{C}}(3,6,50,25)$ is roughly the same. This is expected because both ensembles have the same threshold and approximately the same scaling behavior, given by \eqref{longchain}.

\begin{table}[b]
\begin{center}
\scalebox{0.9}{%
\begin{tabular}{|c|c|}\hline
$p$ & $\pe_{\widetilde{\mathcal{C}}(3,6,50,p)}$  \\\hline
$9$ & $0.467$ \\
$16$ & $0.468$ \\
$22$ & $0.472$ \\
$24$ & $0.478$ \\
$25$ & $0.488$ \\
\hline
\end{tabular}}
\end{center}
\caption{BP thresholds of the $\widetilde{\mathcal{C}}(3,6,50,p)$ code ensembles as a function of $p$.}\label{TABLE1}
\end{table}

\begin{figure*}[h]
\centering\includegraphics[scale=1.5]{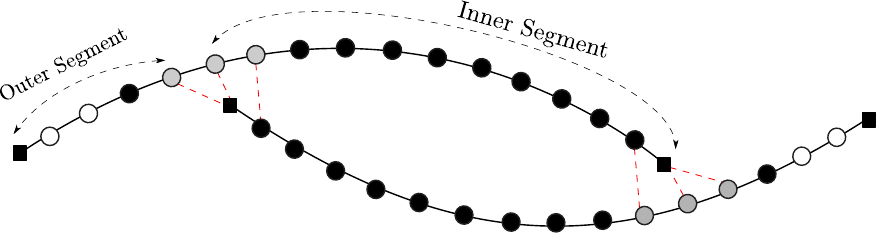}
\caption{Graphical representation of the $\mathcal{L}(3,6,L)$ ensemble for $L=15$. Dashed lines represent the additional edges added,  gray dots indicate positions with variable nodes of degree four, and white dots positions with variable nodes connected to check nodes of degree two or four.}
\label{Fig:loop}
\end{figure*}

In the following section, we  revisit the loop ensemble presented in \cite{Truhachev13, TruhachevICC12}, from the finite-length perspective. The analysis of the  $\widetilde{\mathcal{C}}(3,6,50,p)$ ensemble presented in this section will help us to understand the decoding behavior of the loop ensemble in the large $L$ regime and to propose improved constructions in Section \ref{SectionV}.

\section{The Loop Connected Chain ensemble}\label{loop}

We now turn our attention to a more general connected chain SC-LDPC code ensemble, the loop ensemble \cite{Truhachev13, TruhachevICC12}. We denote it by $\mathcal{L}(l,r,L)$. As before, we use the $(3,6)$-regular LDPC code ensemble to demonstrate behavior; however the techniques and results can easily be extended to general $(l,r)$-regular constructions. The $\mathcal{L}(3,6,L)$ ensemble   consists of two connected $(3,6)$-regular SC-LDPC code chains of length $L$. The low-degree check nodes at one end of the first chain are connected to variable nodes at interior positions of the second chain.  We proceed similarly with the second chain, connecting check nodes at one end of the chain to interior positions in the first chain. The $\mathcal{L}(3,6,L)$ ensemble for $L=15$ is illustrated in Fig. \ref{Fig:loop}. In similar fashion to the construction of the $\widetilde{{C}}(3,6,L,p)$ ensemble,  we use three consecutive positions to connect variable nodes of one chain to check nodes at the end of the other chain, creating variable nodes of degree four at these three positions\footnote{In \cite{Truhachev13}, different ways to connect the chains are compared, and it is shown  that the resulting BP threshold is sensitive to the type of connection for small chain lengths $L$; however, the change for moderate to large $L$ is marginal.}. 
The loop ensemble  can also be  represented by its connectivity matrix:
\begin{align}
\T_{\mathcal{L}(3,6,L)}=\left[
\begin{array}{cc}
\T_{\mathcal{C}(3,6,L)} & \mathbf{L}_2\\
\mathbf{L}_1 & \T_{\mathcal{C}(3,6,L)}
\end{array}
\right],
\end{align}	
where $\mathbf{L}_1$ and $\mathbf{L}_2$ are $(L+2)\times (L+2)$ matrices that define the links between variable nodes in one chain and check nodes in the other chain. The dimension of $\T$ is now $D=2(L+2)$. From (\ref{designrate}), we see that the rates of the $\mathcal{C}(3,6,L)$ and $\mathcal{L}(3,6,L)$ code ensembles are the same.

In~Fig. \ref{Fig:loop},  we observe that there are four regions of the graph where the code is stronger: two regions where variable nodes in two consecutive positions are connected to low-degree check nodes and two regions where there are variable nodes of degree four in three consecutive positions. These four regions divide each chain  into  an ``outer segment'' (from position 1 to the first position containing variable nodes of degree four) and  an ``inner segment'' (from the end of the outer segment to the last position of the chain)  (see Fig. \ref{Fig:loop}). 


\begin{figure}[h]
\centering \includegraphics[scale=0.45]{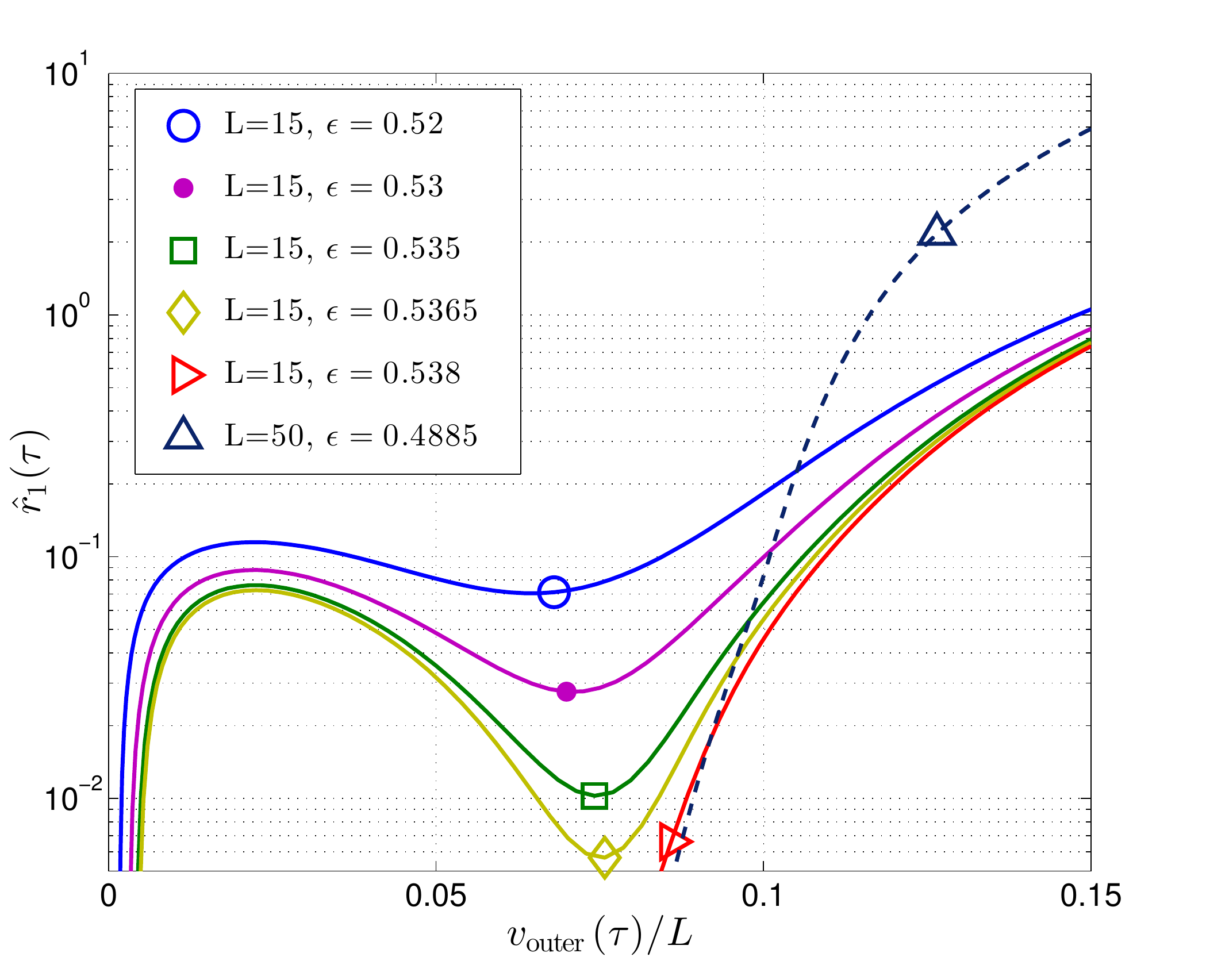}\caption{Expected graph evolution of the normalized number of degree-one check nodes in the graph $\hat{r}_{1}(\tau)$ for the ensemble $\mathcal{L}(3,6,L)$ as a function of $v_{\text{outer}}$ (as defined in \eqref{outer}) for different values of $\epsilon$ above the code threshold.}\label{Fig:outer_segment_r1}

\end{figure}

\begin{table}[h]
\begin{center}
\scalebox{0.9}{%
\begin{tabular}{|c|c|c|c|c|c|}\hline
Rate & Ensemble & $\pe_{\mathcal{L}(3,6,L)}$ & Ensemble & $\pe_{\mathcal{C}(3,6,L)}$ \\\hline
$0.375$ & $\mathcal{L}(3,6,8)$ & $0.5445$ & $\mathcal{C}(3,6,8)$ & $0.522$\\
$0.4$ & $\mathcal{L}(3,6,10)$ & $0.5323$ & $\mathcal{C}(3,6,10)$ & $0.508$\\
$0.4167$ & $\mathcal{L}(3,6,12)$ & $0.5237$ & $\mathcal{C}(3,6,12)$ & $0.495$\\
$0.4333$ & $\mathcal{L}(3,6,15)$ &  $0.5105$ & $\mathcal{C}(3,6,15)$ & $0.489$\\
$0.4444$ & $\mathcal{L}(3,6,18)$ &  $0.4989$ & $\mathcal{C}(3,6,18)$ &  $0.488$ \\
$0.46$ & $\mathcal{L}(3,6,25)$ &  $0.488$ & $\mathcal{C}(3,6,18)$ &  $0.488$ \\
$0.48$ & $\mathcal{L}(3,6,50)$ &  $0.488$ & $\mathcal{C}(3,6,18)$ &  $0.488$ \\
\hline
\end{tabular}}
\end{center}
\caption{BEC thresholds $\epsilon^*$ for several SC-LDPC connected chain loop ensembles $\mathcal{L}(3,6,L)$ and single chain ensembles $\mathcal{C}(3,6,L)$.}\label{tab:thres}
\end{table}

As shown in \cite{Truhachev13, TruhachevICC12}, the asymptotic performance of the $\mathcal{L}(3,6,L)$ ensemble is improved if the sides of the inner loop each have length $\lfloor 2L/3 \rfloor$ and  the maximum check node degree is limited to $6$, \emph{i.e.}, variable nodes in one chain can only be connected to low degree check nodes in the other chain. These requirements are fulfilled if
$\mathbf{L}_1$ has ones at positions  $(1,\ceil{L/3})$, $(1,\ceil{L/3}+1)$, and $(2,\ceil{L/3}+2)$ and $\mathbf{L}_2$  has ones at positions $(L+1,\lfloor 2L/3 \rfloor + 1)$, $(L+2,\lfloor 2L/3 \rfloor + 2)$, and $(L+2,\lfloor 2L/3 \rfloor + 3)$. 
The BEC BP thresholds $\epsilon_{\mathcal{L}(3,6,L)}$ 
for different values of $L$ are shown in Table~\ref{tab:thres}. The thresholds $\epsilon_{\C(3,6,L)}$ of the corresponding single chain SC-LDPC code ensembles of the same rate are given for comparison. It is observed that the thresholds of the connected chain ensembles are  larger than the thresholds of the corresponding single chain ensembles for chain lengths up to $L=18$.

An intuitive understanding of these results is as follows.  For $L\leq18$, the outer segments are quite short and, in addition, they have strong sub-codes at both ends. Hence, compared to the inner segments, which have double the length ($\lfloor 2L/3\rfloor$ positions) and are only connected to variables of degree four, the outer segments are much better protected. Even when we are operating above the  threshold, information encoded in the outer segments can be successfully recovered. For example, in Fig.~\ref{Fig:outer_segment_r1}  we plot the expected graph evolution of the normalized number of degree-one check nodes $\hat{r}_{1}(\tau)$ for the $\mathcal{L}(3,6,15)$ ensemble for different $\epsilon$ values above its BP threshold. Here, we plot  $\hat{r}_1(\tau)$ against $v_{\mathrm{outer}}(\tau)/L$, where
\begin{align}\label{outer}
\hat{v}_{\mathrm{outer}}(\tau)\doteq\sum_{i\in\substack{\mathrm{outer}\\\mathrm{segments}}}\hat{v}_i(\tau)
\end{align}
is the expected fraction of undecoded bits at time $\tau$ that belong to the outer segments. We also include, for comparison, one curve corresponding to the $\mathcal{L}(3,6,50)$ ensemble. We observe that the outer segments can be decoded at channel erasure probabilities up to $\epsilon=0.5365$, while the ensemble threshold is only $\pe_{\mathcal{L}(3,6,15)}=0.5105$. Therefore, for $\pe\leq\pe_{\mathcal{L}(3,6,15)}$, the outer segments are almost surely decoded and, after reducing the graph accordingly, the two inner segments get essentially disconnected and low-degree check nodes are created at their ends.  Consequently, the loop ensemble is decoded as two independent chains of length $2L/3$. Indeed, from Table~\ref{tab:thres} we observe that  the loop thresholds $\pe_{\mathcal{L}(3,6,18)}$,  $\pe_{\mathcal{L}(3,6,15)}$, and $\pe_{\mathcal{L}(3,6,12)}$ are only slightly larger than the thresholds  of the single chain ensembles $\pe_{\mathcal{C}(3,6,12)}$, $\pe_{\mathcal{C}(3,6,10)}$, and  $\pe_{\mathcal{C}(3,6,8)}$, respectively,  with length $2L/3$.

\begin{figure}[h]
\centering \includegraphics[scale=0.45]{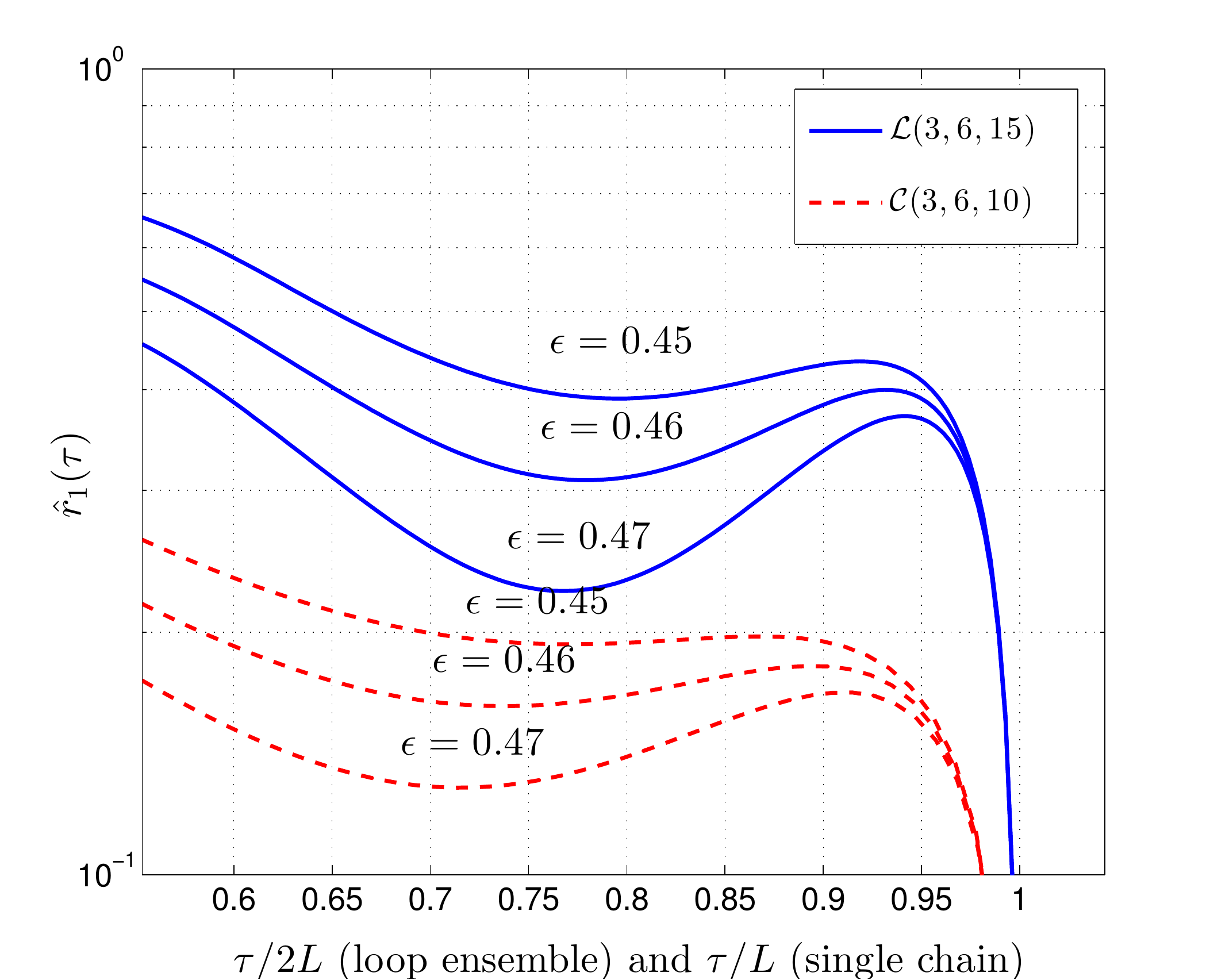}
\caption{Expected graph evolution of the fraction of degree one check nodes $\hat{r}_{1}(\tau)$ for the ensembles $\mathcal{L}(3,6,15)$ and $\mathcal{C}(3,6,10)$ with different  values of $\epsilon$.}\label{Fig:loop_short_r1}
\end{figure}

From the finite-length perspective, once the outer segments are decoded, the decoding of  the two remaining  shorter chains is described by the singe critical-critical point model in \eqref{shortchain}, which is consistent with the  singe critical-critical point behavior observed as well for the $\L(3,6,L)$ ensemble in \cite{Olmos13-2}. For instance, in 
%
%
%
Fig.~\ref{Fig:loop_short_r1} we give expected graph evolution $\hat{r}_1(\tau)$ computed the  for the $\mathcal{L}(3,6,L)$ ensemble with $L=15$ and different  values of $\pe$, where we have also included the corresponding  curves for the $\mathcal{C}(3,6,2L/3=10)$ ensemble\footnote{ To assist visualization  of the results, we have normalized the scaling of  the horizontal axis by $2L$ in the loop case and by $L$ in the single chain case.}.  \textcolor{black}{For both ensembles we can observe a single critical point,  but the  evolution of the $r_1(\tau)$ process for the loop ensemble combines the contribution of the  two connected chains. Therefore, its mean $\hat{r}_1(\tau)$ in
 Fig.~\ref{Fig:loop_short_r1}, but  also its variance, are higher than the corresponding  mean and variance  of the process $r_1(\tau)$ for the  $\mathcal{C}(3,6,10)$  ensemble. By simulation, we have observed that the ratio of the mean $\hat{r}_1(\tau)$ to its standard deviation is approximately the same for both ensembles.}

\begin{figure}[h]
\centering \includegraphics[scale=0.45]{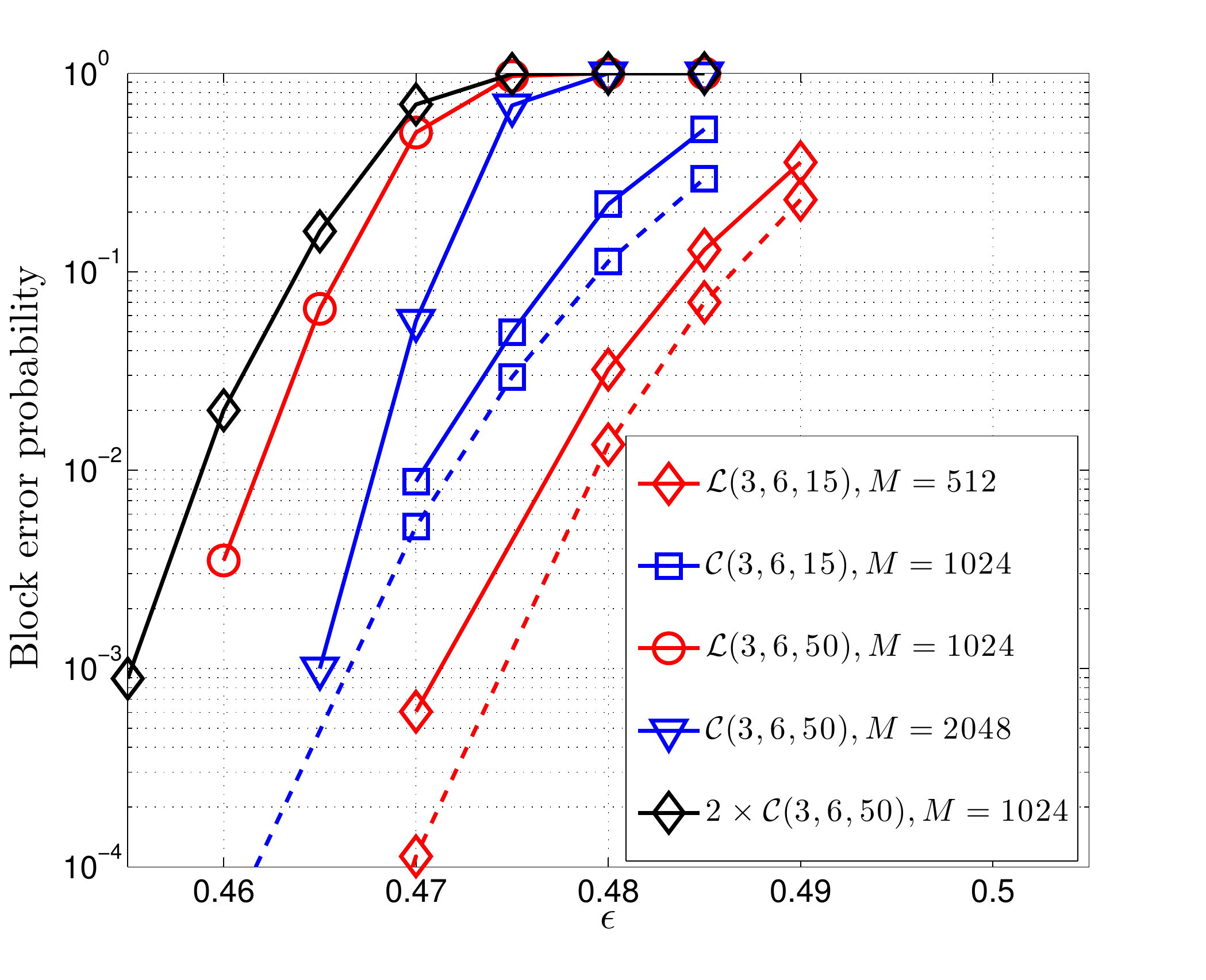}
\caption{Simulated block error rate performance  for the ensembles $\mathcal{L}(3,6,L)$ with $512$ bits per position and $\mathcal{C}(3,6,L)$ with $1024$ bits per position for $L=15$ and $L=50$. In the $L=15$ case, the performance approximation (dashed lines) obtained using the scaling law in \eqref{shortchain} is included.}\label{fig_performance_loop}
\end{figure}

In conclusion, for short chain lengths $L=5,\ldots,18$, the $\mathcal{L}(3,6,L)$ loop ensemble has significantly better thresholds than the $\mathcal{C}(3,6,L)$ single chain ensemble, while maintaining similar finite-length scaling behavior, i.e., the single critical point model. These results explain the better finite-length block error performance of the loop ensemble compared to the single chain ensemble of the same length and rate. More specifically,  both ensembles have the same chain length $L$, but the single chain has $2M$ bits per position compared to $M$ bits per position in the case of the loop.  This is demonstrated by simulation in Fig.~\ref{fig_performance_loop} for the case $L=15$, were we give simulated block error rate curves  for the ensembles $\mathcal{L}(3,6,15)$ with $512$ bits per position and $\mathcal{C}(3,6,15)$ with $1024$ bits per position.  In this figure, we also include the respective performance estimates  for the two ensembles using the scaling law in \eqref{shortchain}.  

\textcolor{black}{The picture, however, is reversed when we consider the loop ensemble with large $L$ values. For instance, in Fig. ~\ref{fig_performance_loop}, the loop ensemble block error rate with $L=50$ is clearly outperformed by the single chain ensemble of the same length and rate.} For large $L$ values, the (longer) outer segments do not display such strong performance above the ensemble threshold. For example, in Fig. \ref{Fig:outer_segment_r1} we include  the $\hat{r}_{1}(\tau) \text{ vs. } v_{\mathrm{outer}}(\tau)/L$ curve for the $\mathcal{L}(3,6,50)$ ensemble and $\pe$ slightly above the code threshold $\pe_{\mathcal{L}(3,6,50)}\approx 0.48815$. Observe that the curve goes to zero for $ v_{\mathrm{outer}}(\tau)>0$, which indicates that, by the time decoding fails, there is still a significant fraction of undecoded bits in the outer segment. For large values of $L$, the two regions in the graph with variable nodes of degree four are far apart from each other and from the two strong boundary positions with low degree check nodes. As demonstrated in Section~\ref{modified_chain} for the $\widetilde{\mathcal{C}}(3,6,50,p)$ ensemble, when isolated, these regions are only effectively decoded for $\epsilon$ values much below  $\pe_{\mathcal{L}(3,6,L)}$. In practice, this means a)  there is no gain in threshold with respect to the $\mathcal{C}(3,6,L)$ ensemble and b) that PD for the $\L(3,6,L)$ ensemble is performed in a wave-like fashion from the boundary positions with low-degree check nodes towards the interior positions.  Therefore, close to the threshold, there is no  practical advantage with  respect to an SC-LDPC code   consisting of two independent long chains of length $L$ and $M$ bits per positions. Using \eqref{longchain}, the block error probability of this ensemble can be estimated as follows:
\begin{align}\label{longchain2}
P_{2\times\mathcal{C}(l,r,L)}&\approx 1-\exp\left(-2\frac{\epsilon L \overline{y}}{\mu_0(M,\epsilon,l,r)}\right)
\end{align}
and this error probability is much worse than for a single chain of the same rate, length $L$, and $2M$ bits per position \cite{Olmos13-2,Olmos13-3}, since $\mu_0$ in \eqref{longchain} grows exponentially in $M$. We demonstrate these results in Fig. ~\ref{fig_performance_loop}, where we give simulated block error rates for the $\mathcal{L}(3,6,50)$ ensemble with $1024$ bits per position and for the $\mathcal{C}(3,6,50)$ ensemble with $2048$ bits per position. In addition, to show that, close to the threshold, the performance of the $\mathcal{L}(3,6,50)$ ensemble scales approximately as in \eqref{longchain2}, we also include the simulated block error performance of a  code consisting of two independent $\C(3,6,50)$ chains with $M=1024$ bits per position $(\Diamond)$. 

To summarize, while we have seen that  significant performance gains  can be obtained by connecting short chains, for long chains it is much better to increase the number of bits $M$ per position in a single SC-LDPC chain rather than trying to obtain improvements by connecting multiple chains. Nevertheless, we will show in the next section that this result does not prevent us from finding  SC-LDPC codes formed by connecting long chains that lead to substantial performance improvements.

%
%

\section{A different approach: connecting consecutive chains}\label{SectionV}

We start by considering the encoding and transmission of codewords using a  single $\mathcal{C}(3,6,L)$ code chain.  The information stream is divided into blocks of $ML\mathtt{r}_{\mathcal{C}(3,6,L)}$ bits, which are then independently encoded, transmitted, and decoded at the receiver. In the following, by transmitting independent chains we mean the transmission of independent  codewords that belong to a  $\mathcal{C}(3,6,L)$ code chain. From \eqref{longchain}, the probability that there is at least one decoding failure in  $N$ independent consecutive chains  scales with $M$ and $L$ as:
\begin{align}\label{longchainN}
P_{N\times\mathcal{C}(l,r,L)}&\approx 1-\exp\left(-N\frac{\epsilon L \overline{y}}{\mu_0(M,\epsilon,l,r)}\right).
\end{align}

In this section, we show that the performance of this system can be  improved if we do not transmit independent chains, but instead data is encoded in a continuous fashion using a convolutional-like structure based on connected SC-LDPC code chains. We refer to such an encoding scheme  as continuous chain (CC) transmission of SC-LDPC codes. More specifically, $N$ consecutive SC-LDPC code chains, that without CC transmission are independently transmitted and decoded, will be now connected such that  the overall robustness against failures is improved and the decoding complexity and decoding delay are not significantly affected. 

In the following, we present some examples that illustrate the benefits of the CC coding/transmission scheme.  We first propose some SC-LDPC code ensembles specifically designed for this application and then we discuss the feasibility of CC transmission.



\subsection{SC-LDPC code ensembles for CC transmission}\label{CCTcodes}

In previous sections we have seen that, when connecting SC-LDPC code chains, the local degree distribution around the connecting points plays an crucial role.
In particular, the analysis of the $\widetilde{\mathcal{C}}(3,6,L,p)$ ensemble in Section \ref{modified_chain}  showed that the two regions formed by variables of degree four result in a decoding bottleneck unless they are located very close together, thus forming a single   well protected graph region. We drew similar conclusions when analyzing the loop ensemble $\L(l,r,L)$ for large $L$, where the distance between connection points is so large that they do not really help to improve the decoding performance compared to an ensemble consisting of two single chains. In the light of these results, we now propose alternative structures based on connecting long chains that are  suited for CC transmission. The main design difference is that  we now break the symmetry of the loop ensemble 
by connecting two  single chains such that one is  more robust against failures while  the other maintains approximately the same decoding properties as a single chain. 
%
%
%
%

\begin{figure}[h]
\centering\includegraphics[scale=1.5]{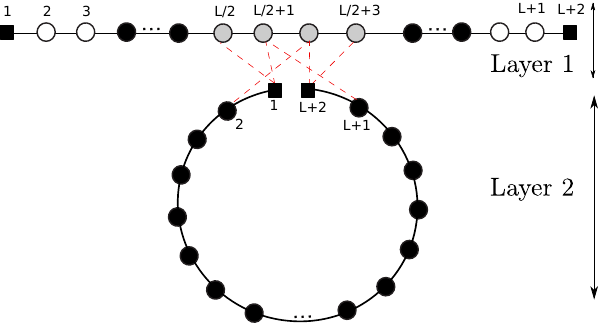}
\caption{Graphical representation of the  $\mathcal{S}(3,6,L,2)$ ensemble in \eqref{eqCCT1}.}
\label{Fig:CCT1}
\end{figure}

This is precisely the effect achieved by the ensemble illustrated in Fig. \ref{Fig:CCT1}. This ensemble consists of two layers with one  single chain of length $L$ per layer, and it is denoted by $\mathcal{S}(3,6,L,2)$. In general, we denote such ensembles as $\mathcal{S}(3,6,L,N)$, where $N$  stands for the number of layers. The layer $j$ chain  is simply referred to as chain $j$.
In Fig. \ref{Fig:CCT1}, the chains in the two layers have been connected to increase the protection of the variable nodes in the four  middle positions of  chain  1, from position  $\ceil{L/2}$ to position $\ceil{L/2}+3$, 
by adding the following edges: 
\begin{itemize}
\item One edge  from each variable node at position $\ceil{L/2}$ in chain 1 to a check node at position $1$ in chain 2.
\item Two edges from each variable node at position $\ceil{L/2}+1$ in  chain $1$ to check nodes at positions $1$ and $(L+1)$ in chain $2$.
\item Two edges from each variable node at position $\ceil{L/2}+2$ in chain $1$ to check nodes at positions $2$ and $(L+2)$ in chain $2$.
\item One edge from each variable node at position $\ceil{L/2}+3$ in chain $1$ to a check node at position $(L+2)$ in chain 2.
\end{itemize}
The connectivity matrix of this ensemble is given by
\begin{align}\label{eqCCT1}
\T_{\mathcal{S}(3,6,L,2)}=\left[
\begin{array}{cc}
\T_{\mathcal{C}(3,6,L)} & \mathbf{0}\\
\mathbf{CCT} & \T_{\mathcal{C}(3,6,L)}
\end{array}
\right],
\end{align}
where $\mathbf{CCT}$ is an $(L+2)\times (L+2)$ matrix with ones at positions $(1, \ceil{L/2})$, $(1,\ceil{L/2}+1)$, and $(2,\ceil{L/2}+2)$, which corresponds to the connections between the left end of the second chain and the middle positions of the first chain, and at positions $(L+1,\ceil{L/2}+1)$, $(L+2,\ceil{L/2}+2)$, and $(L+2,\ceil{L/2}+3)$, which corresponds to the connections between the right end of the second chain and the middle positions of the first chain.  In this way, we create a region with much better protection in the middle of  chain 1, where the variable  degree profile is $(4,5,5,4)$, and all check nodes in chain 2 now have degree $6$.
Note that, while chain 1 contains three strong sub-codes, all variable nodes in chain 2 are of degree 3 and all check nodes are of degree 6. Therefore, to be effectively decoded, chain 2 relies on the decoding of the four middle positions of chain 1, where the variable nodes are better protected. Note that, if positions $\ceil{L/2},\ldots,\ceil{L/2}+3$ in chain 1 are correctly decoded, then   chain 2 has  the same DD as the  SC-LDPC single chain discussed in Section \ref{singlechain}.  Decoding chain 2 is  then similar to  decoding  the  $\widetilde{\mathcal{C}}(3,6,L,\ceil{L/2})$ ensemble discussed in Section \ref{modified_chain}, i.e., to initiate wave-like decoding towards the interior positions, decoding relies  on the decoding of a single protected graph region where all check nodes  are of degree $6$ and the variable nodes are of degrees $(4,5,5,4)$.

%

By computing the expected evolution of degree-one check nodes in the graph for the $\mathcal{S}(3,6,L,2)$ ensemble, we observe that, despite the fact that the  $\mathcal{S}(3,6,L,2)$ threshold is exactly the same as for the single SC-LDPC code chain $(\pe_{\mathcal{S}(3,6,L,2)}\approx 0.48815)$, all variable nodes in positions $(\ceil{L/2},\ldots,\ceil{L/2}+3)$ in  chain 1 can be successfully decoded up to an erasure probability of $\epsilon=0.505$. Therefore, for $\pe\leq\pe_{\mathcal{S}(3,6,L,2)}$, these positions are almost surely decoded, which has two important consequences. First,  the finite-length error rate in chain 2 will be similar to that of the single SC-LDPC code chain and, second,  low-degree check nodes will be created at positions $\ceil{L/2}-1,\ceil{L/2},\ceil{L/2}+3$, and $\ceil{L/2}+4$ in chain 1. Consequently, this chain is essentially decoded as  two interdependent chains of length $\ceil{L/2}$. Further, if we use intermediate chain lengths, \emph{i.e.}, $L\in[40,70]$, the decoding of each of the two segments of length $\ceil{L/2}$ is well predicted by the single-critical point model in \eqref{shortchain}, and we can benefit from much 
better properties in terms of finite-length scaling behavior, as discussed in Section \ref{scaling}. As a consequence, the block error rate of chain  1 will be better than for a single  SC-LDPC code chain of length $L$. These statements will be further confirmed by simulation in Section \ref{sims}. 

\begin{figure}[h]
\centering\includegraphics[scale=0.5]{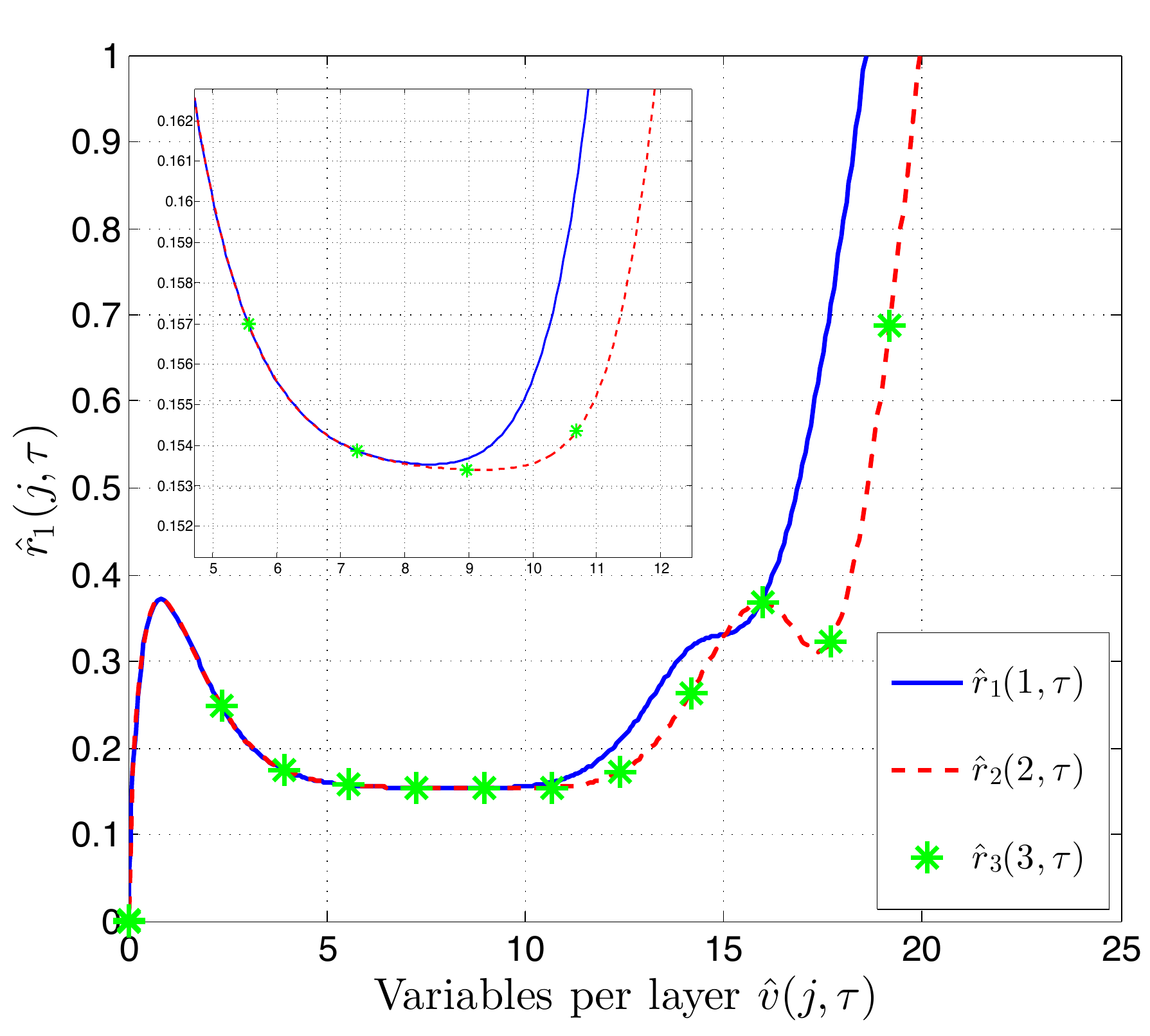}
\caption{Evolution of the normalized number of degree-one check nodes per layer $\hat{r}_{1}(j,\tau)$, $j=1,2,3$, as a function of the  normalized number of variable nodes per layer $\hat{v}(j,\tau)$ during PD for the $\mathcal{S}(3,6,L,4)$ ensemble with $\epsilon=0.45$.}
\label{Fig_N_4}
\end{figure}

\begin{figure*}[h]
\centering \includegraphics[scale=1.5]{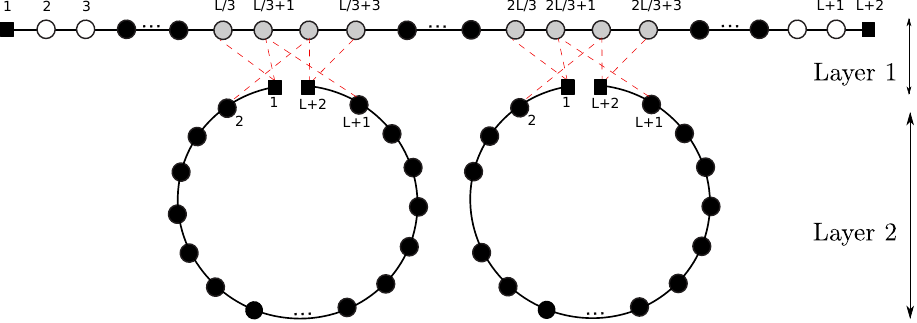}\\
\caption{Graphical representation of the  $\mathcal{S}(3,6,L,2,2)$ ensemble.}
\label{CCT3}
\end{figure*}
%


The design methodology used to construct the $\mathcal{S}(3,6,L,2)$ ensemble can be extended to an arbitrary number  $N$ of layers, and the resulting ensemble has the same rate as the  $\mathcal{C}(3,6,L)$ ensemble for any number of layers.  For instance, the connectivity matrix of the  $\mathcal{S}(3,6,L,3)$ ensemble is given by 
\begin{align}\label{eqCCT2}
\T_{\mathcal{S}(3,6,L,3)}=\left[
\begin{array}{ccc}
\T_{\mathcal{C}(3,6,L)} & \mathbf{0} & \mathbf{0}\\
\mathbf{CCT} & \T_{\mathcal{C}(3,6,L)} & \mathbf{0}\\
\mathbf{0} & \mathbf{CCT}  & \T_{\mathcal{C}(3,6,L)} 
\end{array}
\right],
\end{align}
where the $\mathbf{CCT} $ matrix is of the same form as in \eqref{eqCCT1}. By using the low-degree check nodes of  chain 3,  we increase the protection of the variable nodes at  positions  $\ceil{L/2},\ldots,\ceil{L/2}+3$ in  chain 2, which yields a performance improvement  similar to that obtained in chain 1. More specifically, once the  positions in the middle of the chain 1 have been decoded, besides splitting this chain into two sub-chains of length $\ceil{L/2}$,  low-degree check nodes are created at the boundary positions of   chain 2. This chain now looks similar to chain 1 at the beginning of the decoding process and, consequently, the  performance in layers 1 and 2 is approximately the same. \textcolor{black}{Alternatively, for an arbitrary number $N$ of layers, the finite length performance per chain can be analyzed by computing the $\mathcal{S}(3,6,L,N)$ mean graph evolution during PD, as discussed in Section \ref{expected}. For $N=4$,  in Fig. \ref{Fig_N_4} we plot the expected graph evolution of the normalized number of degree-one check nodes per layer $\hat{r}_{1}(j,\tau)$, $j=1,\ldots,N-1$, as a function of the normalized number of variable nodes per layer $\hat{v}(j,\tau)$. As we can observe from the magnification in the left upper corner, the finite-length performance of chains 1 to 3 is determined by a single critical point of approximately the same height, which indicates a similar exposure to decoding failures.  We will confirm this result in Section \ref{sims}.}

At this point, it is important to clarify that, even though $N$ consecutive chains are now linked in the $\mathcal{S}(3,6,L,N)$ ensemble, a decoding failure in a particular chain does not prevent the rest of chains from being successfully decoded. As discussed above for the $\mathcal{S}(3,6,L,2)$ ensemble, the highly protected region created in the middle of each chain  is decoded almost surely at any erasure below the $\mathcal{S}(3,6,L,N)$ threshold, and this means that decoding failures at chain $j$ do not propagate to lower chains. Therefore, the ratio  of the number of strongly protected chains to the total number of chains is given by $\eta_{\mathcal{S}(3,6,L,N)}=\frac{N-1}{N}$.

The key idea behind the performance improvement experienced by the different chains is to break them into chains of shorter lengths for which the finite-length scaling behavior is more robust. When the rate constraints are even stricter and, as a consequence, we must use even longer chains, \emph{e.g.}, $L=100$, the  $\mathcal{S}(3,6,L,N)$ ensemble  might not be able to provide the same gain per layer as for a shorter chain, \emph{e.g.}, $L=50$. 
In this case,  we can break each chain of length $L=100$ into three shorter segments  by creating two intermediate regions with stronger protection. We illustrate this construction for a two-layer case in Fig. \ref{CCT3}, and we denote such ensembles by $\mathcal{S}(3,6,L,N,t)$, where $t$ refers to the number of chains in each layer that are used to improve the decoding of each chain in the above layer\footnote{The ensemble $\mathcal{S}(3,6,L,N,t)$ for $t=1$ is the one discussed above and  simply denoted by $\mathcal{S}(3,6,L,N)$.}. In this case $t=2$.

Note that, if we add another layer to the ensemble in Fig. \ref{CCT3}, we will need four chains to improve the performance of the  chains in layers 1 and  2. For an arbitrary number $N$ of layers, the ratio of the number of strongly protected chains to the total number of chains is:
\begin{align}
\eta_{\mathcal{S}(3,6,L,N,t)}=\frac{\displaystyle\sum_{j=1}^{N-1}t^{j-1}}{\displaystyle\sum_{j=1}^{N}t^{j-1}},
\end{align}
which  tends  to $1/t$ with increasing $N$. For the ensemble $\mathcal{S}(3,6,L,N,2)$, this implies that, when using CC transmission, half of the chains will enjoy  better performance with no significant increase in  encoding/decoding complexity, as we explain in the next section.

\subsection{Feasibility of CC  transmission}\label{feasibility}

Without  loss of generality, in this section we consider   CC transmission  using the  $\mathcal{S}(3,6,L,N)$ ensemble described above to show that, compared to transmission of independent chain codewords,  CC transmission using connected SC-LDPC code chains only requires some additional memory   and a different transmission order for the encoded bits. 

\subsubsection{Encoding}
%

For a given code belonging the $\mathcal{C}(3,6,L)$ ensemble,  the encoding process can be implemented sequentially using the syndrome former encoder proposed in \cite{pjs+08}. Let $\mathbf{u}^{(i)}$, $i=2,\ldots,L+1$, be a sub-block of $M\mathtt{r}_{\mathcal{C}(3,6,L)}$  information bits and $\mathbf{v}^{(i)}$ the corresponding sub-block of $M$ encoded bits. Using the syndrome former encoder, to compute $\mathbf{v}^{(i)}$ we only need $\mathbf{u}^{(i)}$ and the previously encoded blocks $\mathbf{v}^{(i-1)},\mathbf{v}^{(i-2)},\ldots,$  corresponding to the positions in the SC-LDPC code chain whose variables nodes are connected to the encoded bits $\mathbf{v}^{(i)}$  by the parity check nodes at position $i$.  For example, according to the first two rows of the connectivity matrix  in \eqref{Tmatrix}, $\mathbf{v}^{(1)}$ can be obtained directly from $\mathbf{u}^{(1)}$ and, given $\mathbf{v}^{(1)}$ and $\mathbf{u}^{(2)}$, we can compute $\mathbf{v}^{(2)}$. 


Using the syndrome former encoder, the process of encoding $N$ consecutive layers of the $\mathcal{S}(3,6,L,N)$ ensemble is essentially equivalent in complexity to encoding $N$ consecutive but independent codewords of the single chain ensemble. The only difference is that, after the first chain  in Fig. \ref{Fig:CCT1} has been encoded, the encoding of the first sub-block of the second chain, viz. $\mathbf{v}^{(1)}_{2}$, requires $\mathbf{u}^{(1)}_{2}$ but also $\mathbf{v}^{(\ceil{L/2})},\mathbf{v}^{(\ceil{L/2}+1)},$ and $ \mathbf{v}^{(\ceil{L/2+2})}$. Similarly, to compute the last encoded sub-block in the second chain, viz. $\mathbf{v}^{(L+1)}_{2}$, we also need $\mathbf{v}^{(\ceil{L/2}+1)},\mathbf{v}^{(\ceil{L/2}+2)},$ and $\mathbf{v}^{(\ceil{L/2+3})}$. Therefore, the encoding process has the same complexity as encoding $N$ independent chain codewords, but we have some additional  memory requirements to store the encoded sub-blocks $\mathbf{v}_{j}^{(\ceil{L/2})},\mathbf{v}_{j}^{(\ceil{L/2}+1)},\mathbf{v}_{j}^{(\ceil{L/2+2})}$, and $\mathbf{v}_{j}^{(\ceil{L/2+3})}$ that are necessary to encode the chain at layer $j+1$, $j=1,\ldots,N-1$.

\vspace{0.5cm}
\subsubsection{Window decoding and transmission order}

Efficient  decoding of long SC-LDPC code chains with low decoding delay is based on windowed BP decoding \cite{lscz10,Iyengar11}. In a nutshell, decoding is restricted to a window of $W$ positions that `slides' over the graph, exploiting the convolutional structure of the SC-LDPC code parity check matrix: as bits in the left most positions of the window are decoded, the window is shifted  right and new bits are included in the decoding window (see \cite{lscz10} and \cite{Iyengar11} for further details). For  sufficiently large $W$, \emph{e.g.}, a window of length $W=12$ positions  for the standard $\mathcal{C}(3,6,L)$  SC-LDPC code chain described in Section \ref{singlechain}, the performance is indistinguishable from a  standard BP decoder, while the delay is much less, since  decoding can be initiated before receiving the entire codeword. 

The same decoding principle can be simply adapted to perform efficient decoding of CC  transmission of SC-LDPC codes. For example, for the $\mathcal{S}(3,6,L,N)$ ensemble with $N=2$ in Fig. \ref{Fig:CCT1}, the window decoder can be initiated at the first position of  chain 1. The window will shift until it reaches the middle positions of the chain, whose bits are better protected since they are also connected to check nodes at the boundary positions of  chain  2. Therefore, to efficiently continue  window decoding of chain 1, channel information from the variable nodes at positions $2$ and $(L+1)$  of  chain  2 must be available.  Note that we need this information even before receiving channel information from  the variable nodes at the remaining positions of  chain  1. Once  chain  1 has been decoded, window decoding of  chain  2 can be started using the information already available from variable node positions $2$ and $L+1$. 

Therefore, implementing efficient window decoding for CC transmission reduces to a change in the order in which the encoded bits are transmitted, so that the receiver can have the necessary information at the appropriate time. Clearly, it is necessary for both the transmitter and the receiver to be aware of the transmission order. Returning to the 2-layer example in Fig. \ref{Fig:CCT1}, it would be sufficient to transmit the encoded blocks $\mathbf{v}^{(i)}_{j}$ for $j=1,2$ and $i=2,\ldots,L+1$ in the following order:
\begin{align}
&\mathbf{v}^{(1)}_{1} \rightarrow  \mathbf{v}^{(2)}_{1} \rightarrow \ldots \rightarrow \mathbf{v}^{(\ceil{L/2}+3)}_{1}  \rightarrow
\mathbf{v}^{(1)}_{2}  \rightarrow  \mathbf{v}^{(L+1)}_{2} \nonumber \\\nonumber\\
&\rightarrow \mathbf{v}^{(\ceil{L/2}+4)}_{1} \rightarrow \ldots  \rightarrow \mathbf{v}^{(L+1)}_{1} \rightarrow \mathbf{v}^{(3)}_{2} \rightarrow \nonumber \ldots \rightarrow \mathbf{v}^{(L)}_{2}.
\end{align}
If we add one layer to the example in Fig. \ref{Fig:CCT1}, \emph{i.e.}, we use the  $\mathcal{S}(3,6,L,N)$ ensemble with $N=3$   for CC transmission, we would use the same transmission policy between the boundary positions  in chain 3, \emph{i.e.}, $\mathbf{v}^{(1)}_{3}$ and  $\mathbf{v}^{(L+1)}_{3}$, and the positions $\mathbf{v}^{(\ceil{L/2}+3)}_{2},\ldots,\mathbf{v}^{(\ceil{L+1})}_{2}$ in chain 2. 



\begin{figure}[h]
\centering \includegraphics[scale=0.45]{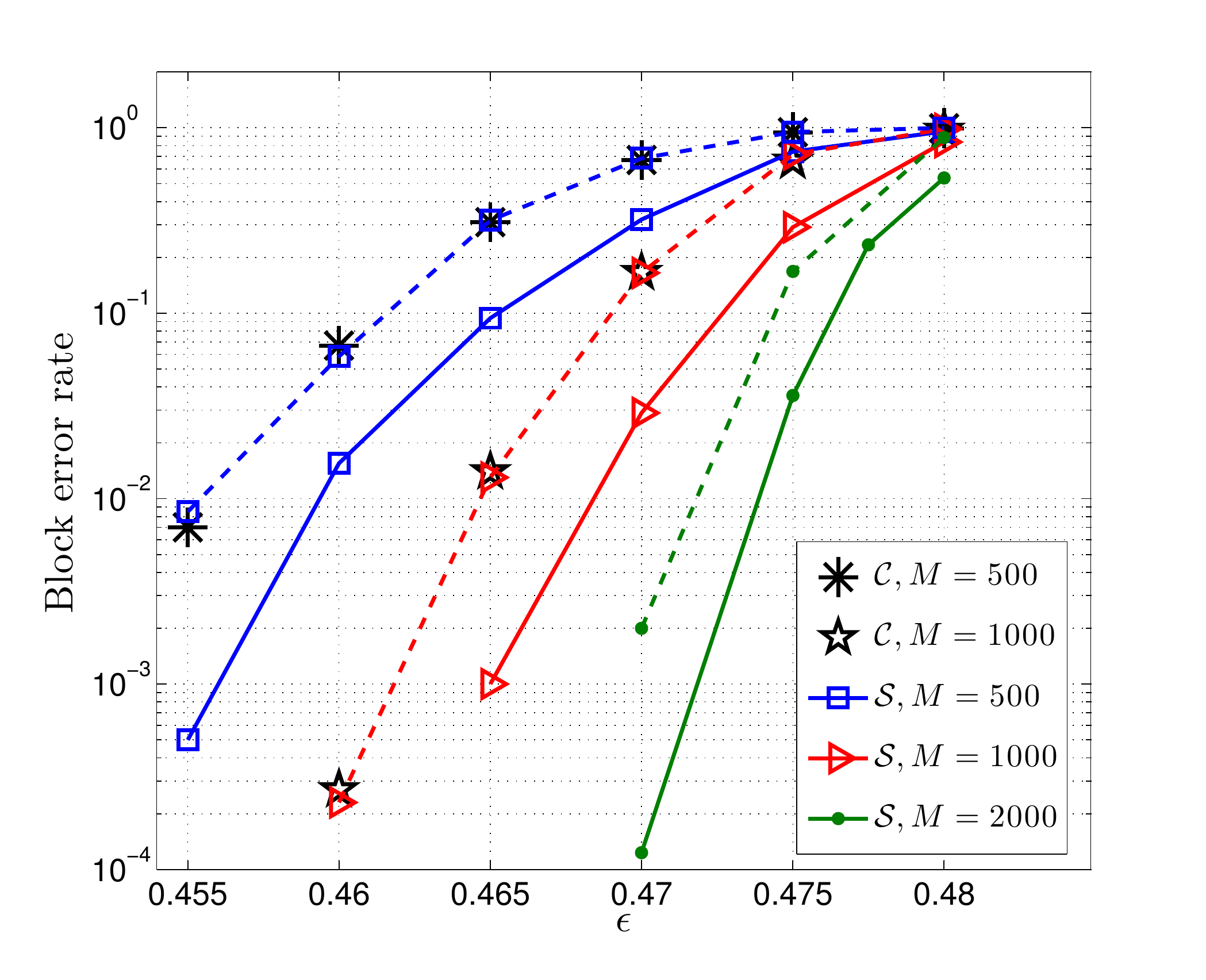}
\caption{Block error rate computed for the chains in layer 1 (solid lines) and layer 2 (dashed lines) of a code drawn from the $\mathcal{S}(3,6,L=50,N=2)$ ensemble. The performance of representative members of the  $\mathcal{C}(3,6,50)$  ensemble is also shown for comparison.}\label{fig_performance_CCT}
\end{figure}

\subsection{Simulation results}\label{sims}

Finally, we include simulation results that corroborate our analysis  of the ensembles presented in Section \ref{CCTcodes} for CC transmission of SC-LDPC codes. The codes are constructed so that they do not contain cycles of length four. Performance curves are obtained after averaging $10^4$ simulations. 

We start by plotting performance results for a code drawn from the $\mathcal{S}(3,6,L=50,N=2)$ ensemble\footnote{All  codes used to generate the simulation results in this section are accessible at the first author's web site:  \tt{ http://www.tsc.uc3m.es/$\sim$olmos/}.} over the BEC. In Fig. \ref{fig_performance_CCT}, we show the block error rate computed for  chain  1 (solid lines) and chain 2 (dashed lines), as well as  the performance of representative members of the ensemble $\mathcal{C}(3,6,50)$. As predicted, we obtain a significant gain in performance for  chain  1, of almost one order of magnitude, even for codes with $2000$ bits per position. Also, we note  that there is no performance degradation in  chain 2 with respect to a single unconnected chain\footnote{\textcolor{black}{Note also that the behavior shown in Fig. \ref{fig_performance_CCT} give us an \emph{unequal error protection capability}, i.e., chain 1 is better protected than chain 2, that can be useful if we have different classes of data.}}. This confirms a) that  our design criterion based on creating a single  powerful connection point is appropriate, and b)  that the modified chain ensemble $\widetilde{\mathcal{C}}(3,6,L,p)$ shown in Fig. \ref{Fig:close_chain} for $p=\ceil{L/2}$ provides the same performance as the standard chain ensemble. 

\begin{figure}[h]
\centering \includegraphics[scale=0.45]{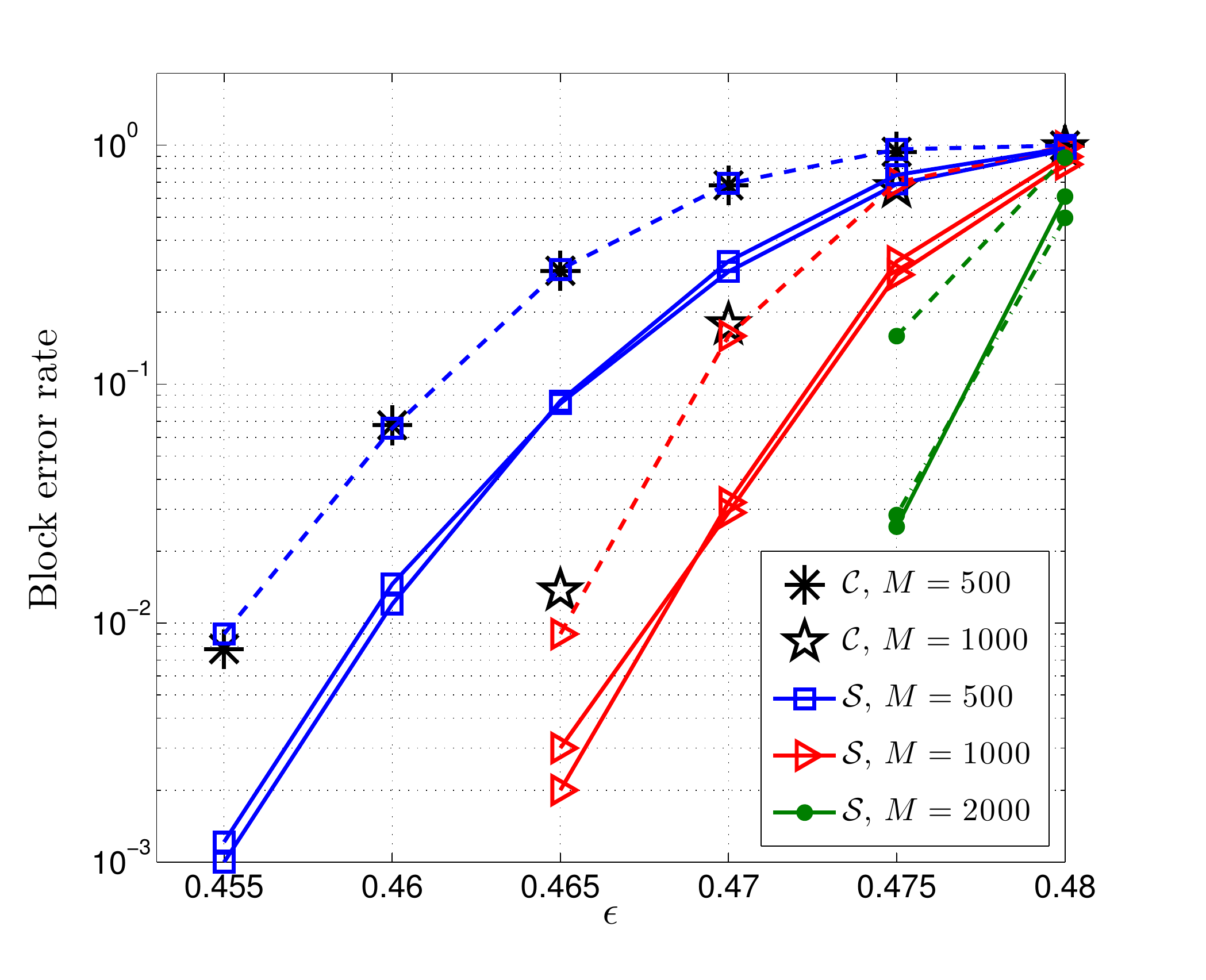}
\caption{Block error rate computed for the chains in layers 1 and 2 (solid lines) and layer 3 (dashed lines) of a code drawn from the $\mathcal{S}(3,6,L=50,N=3)$ ensemble. The performance of representative members of the  $\mathcal{C}(3,6,50)$  ensemble is also shown for comparison.}\label{fig_performance_CCT2}
\end{figure}

In Fig. \ref{fig_performance_CCT2}, we show  results for the ensemble $\mathcal{S}(3,6,50,N)$ with three layers, i.e., $N=3$.  The performance of the first two layers is plotted in solid lines, while the performance of  chain 3 is given in dashed lines.  In addition to what we observed in Fig. \ref{fig_performance_CCT}, note that the performance measured for chains  1 and 2 is essentially the same, and hence there is no degradation in performance in the second layer with respect to the first layer,  even though it does not have low-degree check nodes at each end. This is critical to the success of the CC transmission approach for an arbitrary number of layers, and this behavior has further been confirmed by simulation for ensembles with a larger number of layers. We omit the simulation results for the $\mathcal{S}(3,6,L=100,N=2,t=2)$ ensemble illustrated in Fig. \ref{CCT3}, but, as expected, we have again observed  that the performance of  chain 1 is greatly improved, while the two chains in layer 2 provide similar performance to the SC-LDPC single chain ensemble $\mathcal{C}(3,6,100)$.

%
%

\begin{figure}[h]
\centering \includegraphics[scale=0.45]{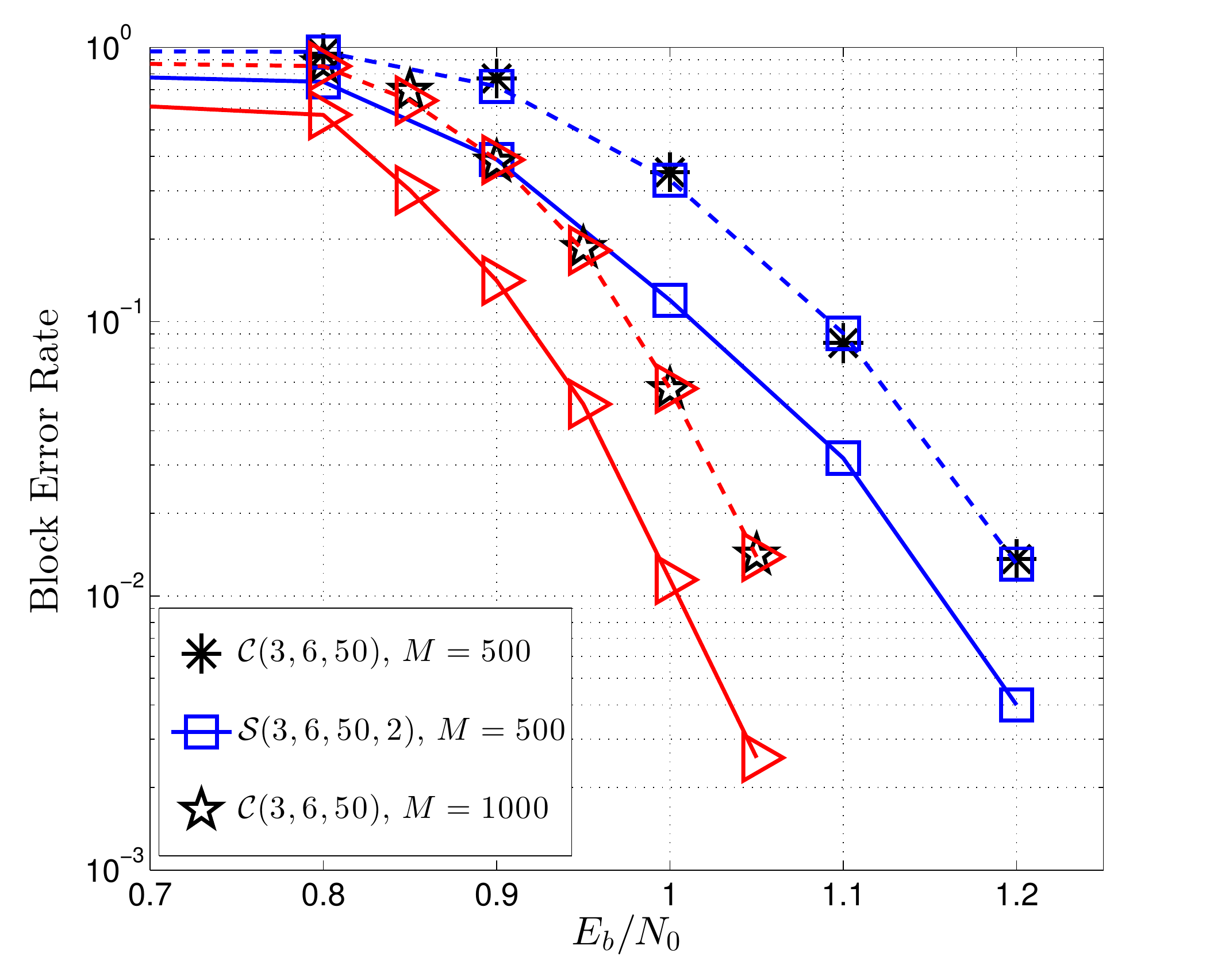}
\caption{Block error rate computed   for the chains in layer 1 (solid lines) and layer 2 (dashed lines) of a code drawn from the $\mathcal{S}(3,6,L=50,N=2)$ ensemble, used for transmission over the BIAWGN channel. The performance of representative members of the  $\mathcal{C}(3,6,50)$  ensemble is also shown for comparison.}\label{BIAWGN}
\end{figure}

\begin{figure}[h]
\centering \includegraphics[scale=0.45]{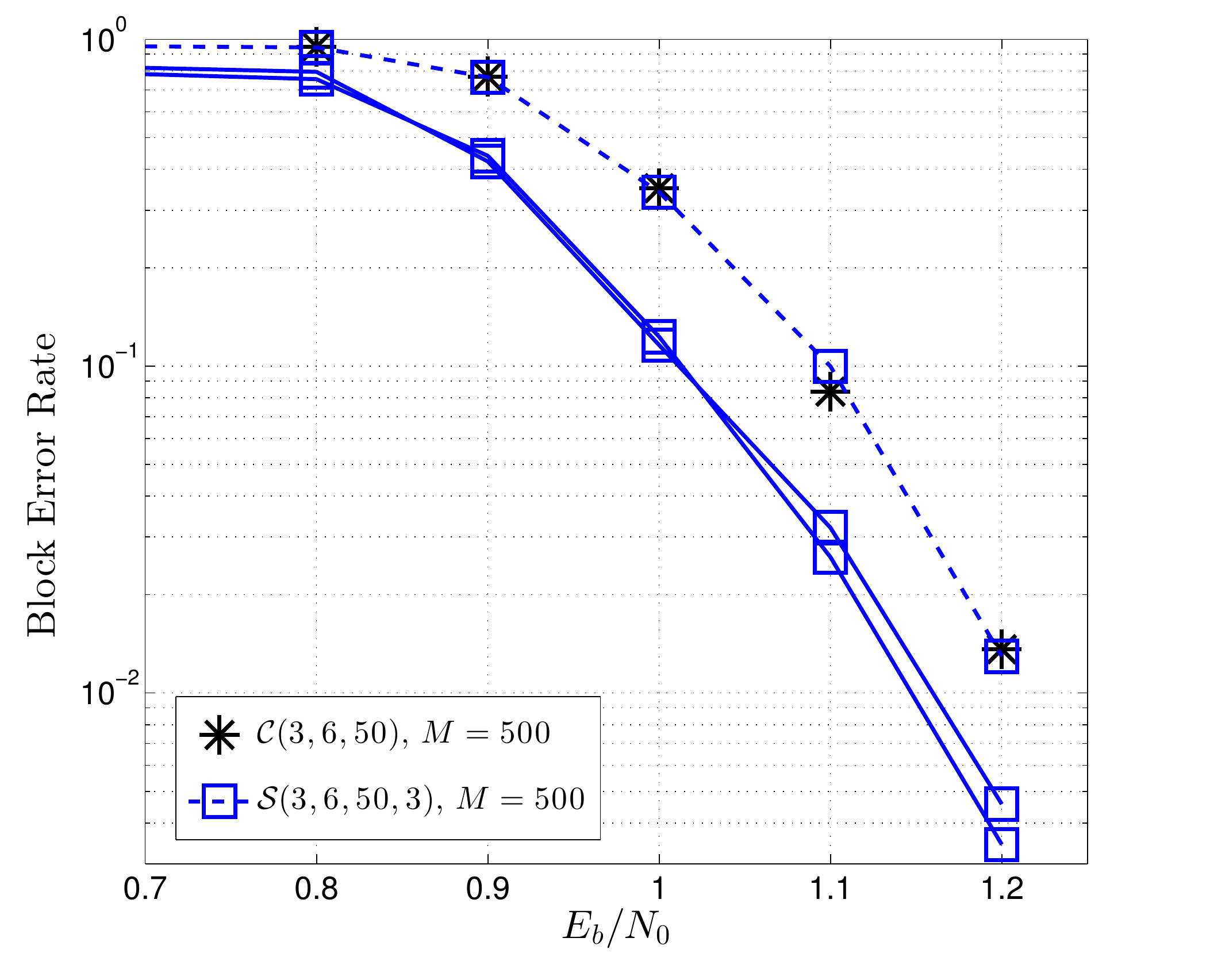}
\caption{Block error rate computed   for the chains in layers 1 and 2 (solid lines) and layer 3 (dashed lines) of a code drawn from the $\mathcal{S}(3,6,L=50,N=3)$ ensemble, used for transmission over the BIAWGN channel. The performance of a representative member of the  $\mathcal{C}(3,6,50)$  ensemble is also shown for comparison.}\label{BIAWGN2}
\end{figure}

Finally, we want to emphasize that, though the  finite-length analysis of the different ensembles was restricted to the BEC, our design and conclusions extend to other channels as well. In Fig. \ref{BIAWGN}, we show results for the $\mathcal{S}(3,6,50,N)$ ensemble with $N=2$  used for transmission over the BIAWGN channel. We represent the performance of chain 2 in dashed lines and  the performance of chain 1 in solid lines. We also include performance results for the single chain ensemble $\mathcal{C}(3,6,50)$. As we can observe, the performance of  chain 1  is substantially improved with respect to chain 2, which in turn has the same performance as the single chain of the same length and rate. In Fig. \ref{BIAWGN2}, we show simulated performance results over the BIAWGN channel for the $\mathcal{S}(3,6,50,N)$ ensemble with $N=3$ and $M=500$ bits per position. The solid lines represent the performance of the chains in the first two layers. As discussed above for the BEC, the performance in the first two layers of the  $\mathcal{S}(3,6,50,3)$ ensemble is not degraded, and we achieve  performance gains with respect a single SC-LDPC code chain. 

\section{Conclusions}\label{future}

In this paper, we have  jointly analyzed  SC-LDPC codes based on a single chain and codes based on connected chains  from a finite-length perspective.  First, we provided a unified vision of existing models, capturing the main scaling behavior between the block error rate and code parameters for a  SC-LDPC code chain. As a novel contribution, we showed that there is an important design aspect regarding the robustness against failures of short  SC-LDPC code chains compared to long chains.  
%
Ensembles based on two connected chains of the same length were proposed in \cite{Truhachev13, TruhachevICC12}, where it was shown they are able to achieve thresholds closer to capacity. In this paper, we have shown that this gain in threshold is a direct consequence of a) the presence of regions in the graph where  variables nodes are better protected, and b) their proximity to one another. Connected chains are shown to be particularly effective when these regions are strong enough to effectively break the chains into shorter, better protected chains. In particular, we have shown the loop ensemble created from two chains of length $L$ essentially behaves, both in threshold and finite-length scaling behavior, as a single chain of shorter length. As a third contribution of the paper,  a novel transmission scheme (CC transmission) designed to boost the performance of a system using long SC-LDPC code chains was introduced. We have shown that by connecting consecutive long chains, rather than transmitting the codewords corresponding to each one independently, we can obtain performance improvements for both the BIAWGN and BEC channels. Moreover, we have shown that CC transmission only requires minor changes to the order in which coded bits are transmitted and some additional memory requirements at the encoder. No significant increase in  encoding/decoding complexity or decoding delay is necessary.


There are many possible variations on the construction of  connected chain SC-LDPC code ensembles presented in this paper. By analyzing  structurally simple ensembles, i.e., random ensembles instead of protograph-based ensembles, we have been able to provide insight into some critical questions regarding  the relation between the structural parameters of connected chain SC-LDPC  codes and their decoding  performance.  Finally, note that the principle of CC transmission is very general and may give rise to other novel encoding/transmission schemes.


\bibliography{prefs.bib,allbib20_03.bib}
\bibliographystyle{IEEEtran}

\appendices
\section{Expected Evolution in One PD Step}\label{A1}
Here we describe the computation of the graph expectations in (\ref{system1}) and (\ref{system2}). Assume the graph at iteration $\ell$ has  DD $\{R_{j,u}(\ell),V_{u}(\ell)\}$ for $u\in[1,D]$ and $j=1,\ldots,r$, and the position where we remove a degree-one check node is denoted as $\mathtt{pos}(\ell)$. At this position, the actual check node removed  is  chosen with uniform probability, and the probability $P(\mathtt{pos}(\ell)=u)\doteq p_u(\ell)$ is given by
\begin{align}
p_u(\ell)=\frac{R_{1,u}(\ell)}{\sum_{i=1}^{D}R_{1,i}(\ell)}.
\end{align}
In the following, let $m=\mathtt{pos}(\ell)$. The variable node connected to the degree-one check node removed is at position $u$ with probability
\begin{align}
\lambda_{m,u}(\ell)=\frac{V_u(\ell)T_{m,u}}{\sum_{i=1}^{D}V_i(\ell)T_{m,i}},
\end{align}
where the denominator represents the total number of variable nodes connected to check nodes at position $m$. Therefore a check node at position $u$ loses one edge with probability
\begin{align}\label{xi}
\xi_{m,u}(\ell)=\sum_{q=1}^{D}T_{u,q}\lambda_{m,q}(\ell).
\end{align}\vspace{-4mm} 

The expected DD evolution at position $m$ is easy to compute, since we just remove an edge of right degree one. Thus $\E[R_{j,m}(\ell+1)-R_{j,m}(\ell)|\mathtt{pos}(\ell)=m]$ is equal to $-1$ for $j=1$ and zero otherwise. For $u\neq m$, the graph  loses one edge with probability $\xi_{m,u}(\ell)$, given by (\ref{xi}). This lost edge was connected to a  degree-$j$ check node with probability
\begin{align}
\displaystyle\frac{R_{j,u}(\ell)}{\sum_{q=1}^{r}R_{q,u}(\ell)}.
\end{align}
In this case, the graph at position $u$ loses $j$ edges of right degree $j$ and gains $j-1$ edges of right degree $j-1$, and the expected graph evolution at position $u\neq m$ is
\begin{align}
\E[R_{j,u}(\ell+1)&-R_{j,u}(\ell)|\mathtt{pos}(\ell)=m]\\\nonumber
&=j\xi_{m,u}(\ell)\frac{R_{j+1,u}(\ell)-R_{j,u}(\ell)}{\sum_{q=1}^{r}R_{q,u}(\ell)},
\end{align}
where for $j=r$, $R_{j+1,u}(\ell)=0$. In addition, we lose a variable node at position $u$ with probability $\lambda_{m,u}(\ell)$. Therefore $\E[V_u(\ell+1)-V_u(\ell)|\mathtt{pos}(\ell)=m]=-\lambda_{m,u}(\ell)$.
Finally, for $2\leq j\leq r$, the expected graph evolution after one iteration of  PD is given by
\begin{align}\label{ExRDD}
&\E[R_{j,u}(\ell+1)-R_{j,u}(\ell)]\\
\nonumber
&=j\sum_{\substack{m=1\\m\neq u}}^{D} \xi_{m,u}(\ell)\frac{R_{j+1,u}(\ell)-R_{j,u}(\ell)}{\sum_{q=1}^{r}R_{q,u}(\ell)}p_m(\ell)\\\nonumber
&=j\left(\frac{R_{j+1,u}(\ell)-R_{j,u}(\ell)}{\sum_{q=1}^{r}R_{q,u}(\ell)}\right)\left(\boldsymbol{p}\boldsymbol{\xi}^{T}_{u}-p_u(\ell)\right),
\end{align}
where $\boldsymbol{p}\doteq\left[p_{1}(\ell) \ldots p_{D}(\ell)\right]$ and $\boldsymbol{\xi}_{u}\doteq[\xi_{1,u},\xi_{2,u}, \ldots,\xi_{D,u}]$.  For the case $j=1$, we only have to include the degree-one check node removed if position $u$ is selected. Thus
\begin{align}
&\E[R_{1,u}(\ell+1)-R_{1,u}(\ell)]=\\\nonumber
&=-p_u(\ell)+\left(\frac{R_{2,u}(\ell)-R_{1,u}(\ell)}{\sum_{q=1}^{r}R_{q,u}(\ell)}\right)\left(\boldsymbol{p}\boldsymbol{\xi}^{T}_{u}-p_u(\ell)\right),
\end{align}
and for the variable nodes we obtain $w\E[V_u(\ell+1)-V_u(\ell)]=-\boldsymbol{p}\boldsymbol{\lambda}_{u}^{T}$,
where $\boldsymbol{\lambda}_{u}\doteq[\lambda_{1,u}, \lambda_{2,u},\ldots,\lambda_{D,u}]$.

\end{document}